\documentstyle[aps,prd, epsf,twocolumn,tighten]{revtex}

\input epsf
\def\plotone#1{\centering \leavevmode
\epsfxsize= 1.0\columnwidth \epsfbox{#1}}

\def\apj{ApJ}
 
\def\mnras{MNRAS}

\def\be{\begin{equation}}
\def\ee{\end{equation}}
\def\bea{\begin{eqnarray}}
\def\eea{\end{eqnarray}}

\def\cmm2{{\,\rm cm^{-2}}}
\def\cm2{{\,{\rm cm}^2}}
\def\cmm3{{\,{\rm cm}^{-3}}}
\def\gcmm3{{\,{\rm g\,cm^{-3}}}}

\def\fun#1#2{\lower3.6pt\vbox{\baselineskip0pt\lineskip.9pt
\ialign{$\mathsurround=0pt#1\hfil##\hfil$\crcr#2\crcr\sim\crcr}}}

\def\eg{{e.g., }}
\def\ie{{i.e., }}
\def\etal{{\it et al. }}

\def\p3m{P$^3$M}

\def\fun#1#2{\lower3.6pt\vbox{\baselineskip0pt\lineskip.9pt
  \ialign{$\mathsurround=0pt#1\hfil##\hfil$\crcr#2\crcr\sim\crcr}}}

\def\spose#1{\hbox to 0pt{#1\hss}}
\def\simlt{\mathrel{\spose{\lower 3pt\hbox{$\mathchar"218$}}
     \raise 2.0pt\hbox{$\mathchar"13C$}}}
\def\simgt{\mathrel{\spose{\lower 3pt\hbox{$\mathchar"218$}}
     \raise 2.0pt\hbox{$\mathchar"13E$}}}
\def \be {\begin{equation}}
\def \en {\end{equation}}
\def \bea {\begin{eqnarray}}
\def \ena {\end{eqnarray}}

\def \bi{\begin{itemize}}
\def \ei{\end{itemize}}

\def \eg {{\it e.g. }}
\def \ie {{\it i.e. }}
\def \etal {{\it et al. }}
\def \l {\ell}

\begin{document}
\twocolumn[\hsize\textwidth\columnwidth\hsize\csname @twocolumnfalse\endcsname
\draft
\title{Elliptical Beams in CMB 
Temperature and Polarization \\
Anisotropy Experiments: An Analytic Approach}
\author{P. Fosalba \and O. Dor\'e \and F.R. Bouchet}
\address{Institut d'Astrophysique de Paris,
98bis, boulevard Arago, F-75014 Paris, France\\
fosalba@iap.fr, dore@iap.fr, bouchet@iap.fr}
\maketitle

\begin{abstract}
We present an analytic approach to the estimation of beam asymmetry
effects in CMB temperature and linear polarization anisotropy experiments.
We derive via perturbative expansions 
simple and accurate results for the case of an 
elliptical Gaussian window. Our results are applied
to investigate the effect of beam ellipticity
in the estimation of full-sky polarization correlation functions
and the covariance matrix of power spectra.
The relevance of this effect is also discussed by forecasting 
errors including beam asymmetry for current and future CMB experiments.
\end{abstract}

\pacs{98.70.Vc}
] 

\section{Introduction}
\label{sec:intro}

As high-resolution CMB experiments explore smaller 
fluctuations in the temperature anisotropy with high sensitivity, 
a better understanding of systematic effects is required to make
more accurate measurements. These systematics have a direct impact on the
ability we have to improve the process of CMB data analysis 
at the level of map-making, power spectrum
estimation and ultimately in constraining cosmological parameters.

A common simplifying assumption in CMB data analysis 
is to take the experimental beam response, i.e, the iso-contours
of constant beam response, to have a perfectly axisymmetric or
{\it circular} shape with a Gaussian profile.
This theoretical approximation introduces systematic errors  
in the statistical analysis at angular scales comparable to the
beam-width, $\sigma$.
Consistently, it bias estimates probing multipole orders
$\l \sim 1/\sigma$ in the spherical harmonic analysis (i.e, the
generalization of flat-space Fourier analysis for full-sky signals) 
of CMB experiments.   
 
As far as the main lobe is concerned, 
experimental beam responses for {\it off-axis} detectors are well-known to 
exhibit {\it asymmetric} shapes very well described by an 
elliptical shape with a Gaussian profile,
as discussed for several experiments in the literature, e.g,  
Planck \cite{BuMa98}, \cite{Maetal00}, Maxima-1 \cite{WuSt00} 
and Python-V \cite{SoRa01}.
However,
the effect of beam asymmetry has been
investigated only recently and the approach taken up to now has relied on
semi-analytic \cite{SoRa01} or full numerical integration 
\cite{BuMa98}, \cite{WuSt00}. 

In this paper we shall introduce an analytic
approach to address the problem of beam asymmetry in CMB experiments.
In particular, we conveniently describe  
an elliptical Gaussian window in terms of a perturbative expansion around a 
circular Gaussian one. As it will be shown below, this description allows 
a simple and intuitive discussion of the beam harmonic transform, 
the full-sky correlation and covariance matrices for both  
total intensity and linear polarization anisotropy observations.     

The paper is organized as follows: in \S\ref{sec:intensity} we present
our analytic approach and derive the spherical harmonic transform of
the total intensity beam. A detailed discussion of the effect of
ellipticity to first order is provided in  \S\ref{sec:slight}.
These results are validated numerically in 
\S\ref{sec:numerics}. Results for linear polarization experiments are 
given in \S\ref{sec:linpol}. We implement this formalism
to calculate full-sky polarization correlation functions in
\S\ref{sec:spectra}. Errors in temperature and polarization power spectra
are discussed in \S\ref{sec:covariance}. Finally, we present a 
general discussion and our main conclusions in \S\ref{sec:disc}.


\section{Beam Spherical Harmonic Transform: Total Intensity}
\label{sec:intensity}

Let us consider the beam response, B, to the total intensity 
sky distribution in a CMB temperature anisotropy experiment. 
For single-dish experiments with high spatial resolution, 
the beam geometry can 
be accurately described in the flat-sky approximation. 
Within this approximation, 
an elliptical Gaussian window function can be expressed in 
cartesian coordinates,
\bea
B(x, y) & = & B_0(\sigma_a,\sigma_b)\ 
\exp\Bigl[-{x^2 \over {2\sigma^2_a}} - {y^2 \over {2\sigma^2_b}} \Bigr]
\label{bflat}
\ena 
where we define $\sigma_a$ and $\sigma_b$ as the beam-widths in the
major (x) and minor (y) axis, and the normalization is given
by $B_0(\sigma_a,\sigma_b) = 1/(2\pi \sigma_a \sigma_b)$.

The Fourier transform of the flat-sky elliptical window is simply
given by,
\bea
B(k_x,k_y) = \exp\Bigl[-{k^2_x \sigma_a^2 \over 2}-{k^2_y \sigma_b^2 \over 2}\Bigr]
\label{bfourier}
\ena
being $k_x$ and $k_y$ the modes along the major and minor axis of the
ellipse, respectively.
However, the Fourier analysis is only accurate for small
patches of the sky (\ie patches covering an area of a few deg$^2$ or smaller).

For full-sky CMB analysis we shall introduce a decomposition
of the window function in the spherical harmonic basis 
$Y_{\l \,m}(\theta,\phi)$,
\be
B(\theta,\phi) = 
\sum_{\l} \sum_{m=-\l}^{+\l} ~b_{\l \,m} \ Y_{\l \,m}(\theta,\phi)
\en
where $\l \approx \pi/\theta$ is the multipole order and 
$b_{\l \,m}$ are the coefficients of the harmonic transform,
\be
b_{\l \,m} =  \int\ d\Omega ~B(\theta,\phi)\ Y^*_{\l \,m}(\theta,\phi) 
\label{blmdef}
\en
being $d\Omega =\sin\theta d\theta d\phi$ the differential solid angle.
Above, we have rewritten the elliptical
window function $B(\theta,\phi)$ in the (planar) polar coordinates, 
$x=\theta\cos(\phi-\omega)$ 
and $y=\theta\sin(\phi-\omega)$, 
\bea
B(\theta, \phi) & = & 
B_0\ \exp\Bigl[-{\theta^2 \over 2\sigma_b^2} f(\phi)\Bigr] 
\label{bpolar}
\ena 
where $f(\phi) \equiv 1 - \chi\cos^2(\phi-\omega)$ describes the
deviations from a circular (or axisymmetric) Gaussian window
and the ellipticity
parameter $\chi \equiv 1-
(\sigma_b/ \sigma_a)^2$, is defined within the range $1 > \chi \ge 0$.
We have introduced an arbitrary phase $\omega$ which defines
the orientation of the major axis of the elliptical beam 
in polar coordinates.  
The circular Gaussian window is thus 
recovered for the limiting case $\chi = 0$.

However, 
the above integral for the spherical harmonic transform of the elliptical
Gaussian window Eq.(\ref{blmdef}) has no exact analytic solution
and one has to resort to semi-analytic approaches or full 
numerical integration to evaluate it (see eg, \cite{SoRa01}).
  
In this section we shall show that Eq.(\ref{blmdef}) can be solved 
analytically by introducing a convenient Taylor expansion
of the elliptical (non-axisymmetric) window around a circular
(axisymmetric) one. 
This perturbative expansion yields
a series in powers of the ellipticity parameter $\chi$,
\bea
B(\theta, \phi) & = & B_0\ \exp\Bigl[-{\theta^2 \over 2\sigma^2} + {\theta^2
\over 2\sigma^2} \chi\cos^2(\phi - \omega)\Bigr] \nonumber \\
& \approx & B_0 \Bigl[B(\theta) + \chi {\theta^2 \over 2\sigma^2}
B(\theta) \cos^2(\phi-\omega)\Bigr]  + {\cal O} (\chi^2)
\label{bpert}
\ena
where the first term corresponds to a circular Gaussian beam 
$B(\theta) = \exp\Bigl[-\theta^2 /2\sigma^2 \Bigr]$ of 
beam-width $\sigma_b$ (the minor axis of the ellipse; we shall denote
$\sigma_b = \sigma$ in what follows for simplicity) and 
$B_0 = [\int d\Omega B(\theta,\phi)]^{-1}$ is the beam normalization.

The above expansion of the window function in real space
Eq.(\ref{bpert}) leads to an analog expansion in harmonic space.
\be
b_{\l \,m} = \sum_{n} b_{\l \,m}^{(n)} {\chi^n \over n!} =
b_{\l \,m}^{(0)} + b_{\l \,m}^{(1)} \chi + {\cal O} (\chi^2)
\label{blmpert}
\en
The $n$-th order term of the harmonic transform can be 
exactly integrated. In particular, only even m-modes have a
non-vanishing harmonic transform
\footnote{This is due to the azimuthal symmetry of the elliptical geometry,
what is realized in the $\cos^2\phi$ factor of $f(\phi)$ 
in Eq.(\ref{bpolar}).} 
which reads (see Appendix 1 for the key steps of the derivation),
\be
b_{\l \,m}^{(n)}  = {2\pi \over 2^{2n-m/2}} ~N_{\l\,-m} ~{{2n!}\over (n+m/2)!} 
~\sigma^{2+m} ~e^{-z} ~L^{(m)}_{n-m/2}(z)
\label{blmn}
\en
being $z=\l^2 \sigma^2/2$, $N_{\l\,m}$ is the normalization of
the spherical harmonics (see Appendix 1) and $L^{(\alpha)}_{\nu}(z)$
denotes the $\nu^{th}$ order Laguerre polynomials of parameter
$\alpha$
(see Eq.(\ref{app:lagpol}) for explicit forms).

Replacing Eq.(\ref{blmn}) into Eq.(\ref{blmpert}) one gets the final 
expression for the harmonic transform of the elliptical beam
\be
b_{\l \,m} = \sigma^{m} ~N^I_{\l\,m}
~e^{-z} ~\sum^{\infty}_{\nu=0} \gamma_{\nu, m} 
~L^{(m)}_{\nu}(z) ~\chi^{\nu+m/2} 
\label{blm}
\en   
where we define $N^I_{\l\,m} = N_{\l\,-m}/\bar{B_0}$, 
$\bar{B_0} = B_0/(2\pi\sigma^2)$, and
$\gamma_{\nu, m}=(2\nu+m)!/(2^{2\nu+3m/2}(\nu+m/2)!(\nu+m)!)$.
Note that the circular Gaussian beam is recovered when $\chi=0$, in
which case, only the $m=0$ contribution is non-zero,
$b_{\l \,m} = ~N_{\l\,0} ~\exp[-\l^2\sigma^2/2] ~\delta_{m,0}$.
Combining the conjugation rule for spherical harmonics,
$b^{\ast}_{\l\,m} = (-1)^m b_{\l\,-m}$ (where non-zero $m$-contributors
are even for an elliptical beam) and the reality condition 
of the beam transform, $b^{\ast}_{\l\,m} = b_{\l\,m}$ one sees that
that {\em both negative and positive modes have the same harmonic
transform}, $b_{\l\,-m} = b_{\l\,m}$.
Therefore, in what follows, we shall assume $m > 0$ without 
loss of generality. 

Eq.(\ref{blm}) is one of the main results of this paper.
This equation demonstrates that 
the leading order correction to the circular Gaussian
window from a given $m$-mode (for $m > 0$)
is of order ${\cal O} (\chi^{m/2})$.
In other words, contributions from higher m-modes to the elliptical window
function can be identified as higher
order corrections to the circular Gaussian window.

For high resolution experiments, $\sigma \ll 1$ rad, 
{\em the elliptical beam harmonic transform 
is dominated by  the axisymmetric or circular contribution
to the window function}, \ie the $m=0$ mode. It is important to
realize that the circular mode does no longer have a Gaussian profile, due to 
ellipticity corrections (see $\nu \ne 0$ terms in Eq.(\ref{blm})).
To leading order in the small $\chi$-expansion 
($\nu=0$ in Eq.(\ref{blm})), contributions from $m>0$ modes
are highly suppressed,
\be
b_{\l \,m} = \gamma_{0, m} ~\chi^{m/2} ~(\sigma \l)^{m} N_{\l\,0} 
~e^{-\l^2\sigma^2/2} \quad ,
\quad m=2,4,6,\ldots
\label{highm}
\en  
Therefore
non-circular (higher-$m$) modes
only have a non-negligible contribution to the harmonic transform 
with respect to the circular ($m=0$) mode when
$\sigma^2 \l^2 \approx 1/\chi$, which is well beyond the peak of the
window function. The peak of the window is determined from 
the leading order contribution to $b_{\l\,0}$ Eq.(\ref{bl01}).
In fact, the peak location and width can also be accurately estimated from 
an \emph{effective} circular Gaussian window of width,
$\sigma_{eff} = \sigma (1+\chi /4)$.
\footnote{Higher-order
corrections in the perturbative expansion Eq.(\ref{blm}) ie, higher-order
terms in $\nu$ and $m$ only modify significantly this definition 
for very large ellipticities $\chi \simeq 1$.} 
\be
\sigma^2 \l_{peak}^2 \simeq (1-\chi/4)/2
\label{lpeak}
\en
since $\chi<1$.
Note also that the window function peaks at increasingly higher
$\l$-multipoles as one considers higher (non-circular) $m$-modes.  
This is in agreement with recent numerical results \cite{SoRa01}
and provides a simple demonstration for them.

\section{First Order Analysis: Slightly Elliptical Beams}
\label{sec:slight}

For most current and future experiments, such as Boomerang
\cite{Boom}, MAXIMA-1 \cite{Max} and Planck \cite{Planck},
the beam is only slightly elliptical, ie, the 
widths of the major ($\sigma_a$) and minor axis ($\sigma$) of the beam 
differ by less than $20 \%$, $1.2 \ge \sigma_a/\sigma \ge 1$ ~($\chi \le 0.3$).

In this limit, a first order ellipticity correction
to the circular Gaussian beam would give an accurate approximation to the
actual beam harmonic transform, which yields for the 
modes $m=0$ and $m=2$,
\be
b_{\l \,0} = N_{\l\,0} ~e^{-\l^2\sigma^2/2} ~\Bigl[1-{\chi\over
4} \l^2\sigma^2 \Bigr] + {\cal O} (\chi^2)
\label{bl01}
\en
\be
b_{\l \,2} = N_{\l\,0} ~{\chi \over 8} ~\l^2\sigma^2 ~e^{-\l^2\sigma^2/2} 
+ {\cal O} (\chi^2)
\label{bl21}
\en
what shows that
\be
b_{\l \,2} = ~{\chi \over 8} ~\l^2\sigma^2 ~b_{\l \,0} + {\cal O} (\chi^2)
\label{bl2r}
\en
From this equation it is straightforward to see that for $\chi \ll 1$ the
leading order contribution from the $m=2$ mode is a few percent
of that from $m=0$ at the peak of the window 
$\sigma^2 \l_{peak}^2 \approx 1/2$ (see Eq.(\ref{lpeak})).

\begin{figure}
\plotone{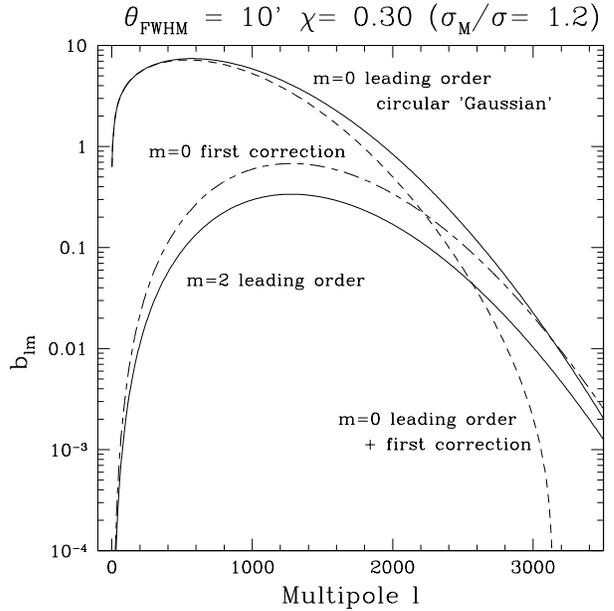}
\caption{Ellipticity corrections to the harmonic transform 
for a slightly elliptical beam, with ellipticity $\chi =0.3$ and
resolution $\theta_{\rm FWHM} = 10^{\prime}$. 
(Solid lines) Leading order terms (for $m=0$ and
$m=2$). The first correction to the circular
($m=0$)
mode (dashed line) 
is of the same order and peaks at the same multipole than the
leading order term for the non-circular ($m=2$) mode (dot-dashed).}
\label{blm_nm}
\end{figure}

However, notice that for the circular mode of the window Eq.(\ref{bl01})
{\em the linear correction 
to the circular Gaussian window} (second term in Eq.(\ref{bl01}))
{\em is of the same order and peaks at the same multipole than the
leading term in the non-circular ($m=2$) mode} (see Eq.(\ref{bl21})).
Therefore, both corrections have to be included to compute the
harmonic transform of the elliptical beam consistently.
This is illustrated in Fig \ref{blm_nm} for an elliptical beam
with $\chi = 0.3 (\sigma_a/\sigma =1.2)$ and $\theta_{\rm FWHM} = 
10^{\prime}$.

Similarly, for highly elliptical beams, higher order $\chi$
corrections to the circular mode become non-negligible and consequently
higher non-circular modes have to be incorporated to calculate
the harmonic transform accurately.
Explicit expresssions for the window function up to second order in the
ellipticity are given in the Appendix A, Eq(\ref{app:blm2or}).
This result arises naturally in the perturbative
approach to the harmonic analysis of
elliptical beams.

\section{Numerical Investigations}
\label{sec:numerics}

In this section we shall 
validate the analytic results 
presented in the previous sections regarding the total intensity window. 
Although the same validation has been carried out for the linear
polarization (see \S\ref{sec:linpol}), we shall concentrate here on the 
total intensity window as the results for polarization are a
straightforward generalization of the total intensity ones.

First of all, we shall test whether the perturbative series 
Eq.(\ref{blmn}) is accurate and how
fast it converges to the numerical solution. 
This analysis is done in \S\ref{sec:converge}.
In addition, we shall see in \S\ref{sec:highm} 
to what extent the {\em scaling solution}
for the higher $m$-modes of the window, Eq.(\ref{highm}), is a good
approximation to the exact solution.
As will be discussed in \S\ref{sec:presc}, prescriptions for an accurate 
computation of the window function from the perturbative solutions
naturally lead to a criteria to
how many higher (non-circular) $m$-modes have to be included in 
a consistent analysis of elliptical windows.

\subsection{Probing the convergence}
\label{sec:converge}


\begin{figure}
\plotone{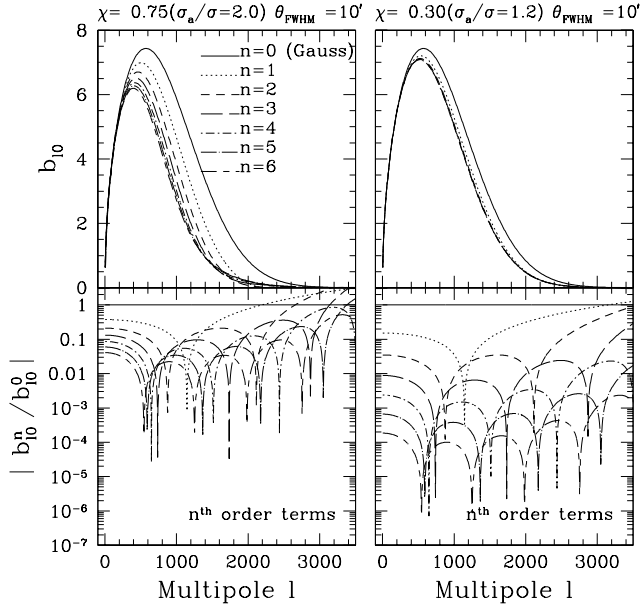} \caption{Probing the convergence. Considering
two
different ellipticities of a given beam of width $\theta_{\rm FWHM} =
10'\displaystyle$, we plot in the upper panel the $n$-th
order expansion of  $b_{\l m}$ and in the lower panel, the
ratio of the $n$-th order correction to the $0$-th one. Both 
plots illustrate the nice and fast convergence of the $\chi$
expansion. \label{fig:plot_blm_n_1}}
\end{figure}

To test our approach, we shall compare the analytic results, Eq.(\ref{blm}),   
to a full numerical
integration of Eq.(\ref{blmdef}) using a Runge-Kutta method of fifth
order \cite{PrTe92}. The fast convergence of the analytical expansion
is illustrated in Fig \ref{fig:plot_blm_n_1}. Indeed, in this figure
we consider for one single beam size,  $\theta_{\rm FWHM} = \sqrt{8\ln
2}\ \sigma =  10'\displaystyle$, two different values of 
the ellipticity parameter, $\chi$~:
$\chi = 0.75\; (\sigma_a/\sigma = 2.0)\displaystyle$ and $\chi =
0.30\; (\sigma_a/\sigma = 1.2)\displaystyle$ shown in the left
and right panels, respectively. 
The upper panels display the expansion $b_{\l\,0}^{(n)}$ for various
$n$. 
In both situations, the convergence is seen
to take place for rather small $n$. 
Comparing this two columns we see also that,
as expected, the
greater the beam ellipticity $\chi$ is, 
the higher the number of terms needed to reach the
convergence. Both this statements will be precised quantitatively below.

The lower panel illustrates this statement by drawing instead the
ratio of the individual $n$-th order terms to the $0$-th order one
(for the same values of $n$ than in the upper panels 
of Fig \ref{fig:plot_blm_n_1}).

\subsection{Higher $m$ modes contribution}
\label{sec:highm}


\begin{figure}
\plotone{ 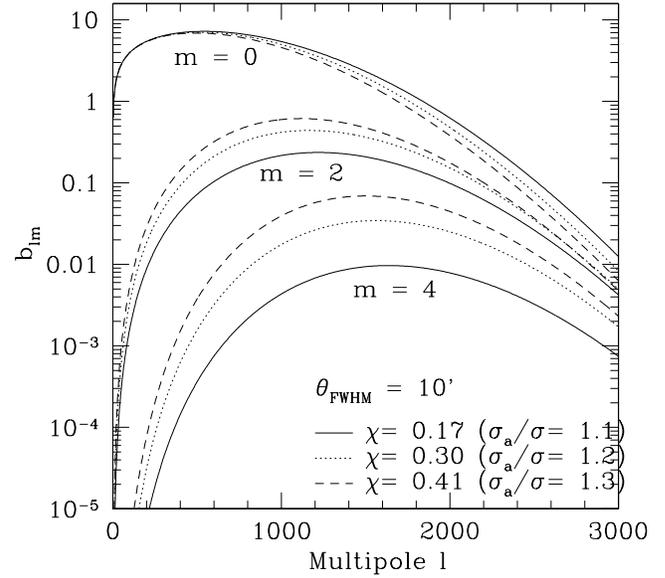} \caption{
Higher-$m$ modes in the elliptical window spherical transform.
\label{fig:plot_blm_m024}}
\end{figure}

An analogous behavior as the one illustrated above for the $m=0$ mode
convergence is seen for $m \ne 0$. We remind here that odd $m$-modes
are null and that only $m \ge 0$ modes are considered since negative modes
have exactly the same harmonic transform, $b_{\l\,m} = b_{\l\,-m}$ 
(see \S\ref{sec:intensity})).
Assuming this convergence, we now
examine the amplitude of the higher $m$ modes contribution, as
they were derived analytically in \S\ref{sec:intensity}. This is
illustrated in fig.(\ref{fig:plot_blm_m024}) for a beam of same width,
\ie $\theta_{\rm FWHM} =
10'\displaystyle$, and for $3$ different values of $\chi$, namely
$\chi = 0.17\; (\sigma_a/\sigma = 1.1)\displaystyle$, $\chi = 0.30\;
(\sigma_a/\sigma = 1.2)\displaystyle$ and $\chi = 0.41\;
(\sigma_a/\sigma = 1.3)\displaystyle$. We plot here the $n$-th order
expansion of $b_{lm}$ for $m=0,2,4$ where $n$ is high enough so that
this expansion is fully converged.

The scalings demonstrated in Eq.(\ref{highm}) are clearly
visible. First we check that $m> 0$  modes amplitude scales as
$(\sigma \l)^{m}$, making them not only sub-dominant (just a few
percent contribution to the beam transform as compared to
circular mode $m=0$) but also shifting
their peak to higher $\l$ as $m$ increases. Second we check also the
scaling with the ellipticity,
$\chi^{m/2}$, that clearly implies that the smaller $\chi$, the more
drastically the $m> 0$ modes are suppressed.

\begin{figure}
\plotone{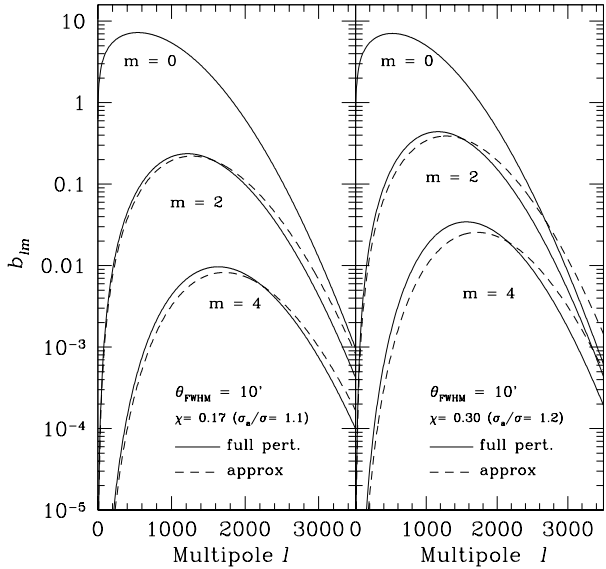} \caption{Probing an approximate
computation of $b_{\l m}$ for $m=2,4$. Considering one beam of width
$\theta_{\rm FWHM} =
10'\displaystyle$, we plot  both the converged expansion of $b_{\l m}$
and an approximate evaluation of it as defined in
Eq.(\ref{highm}). The left panel corresponds to $\chi = 0.17\;
(\sigma_a/\sigma = 1.1)\displaystyle$, and the right one to $\chi =
0.30\;
(\sigma_a/\sigma = 1.2)\displaystyle$. In both situations, the peak
position and
its amplitude are well reproduced. \label{fig:plot_blm_m024_scaling}}
\end{figure}

A direct comparison with the approximate {\em scaling solution} for the
higher $m$-modes Eq.(\ref{highm}) is shown in Fig 
\ref{fig:plot_blm_m024_scaling}.
In particular, the plot shows both the fully
converged expansion of these $m=0,2,4$ modes and the {\em scaling solution}
for rather small values of the ellipticity $\chi$. 
We see that both the peak position and the
amplitude are pretty well reproduced. 
Therefore  the scaling solution, Eq(\ref{highm}),
is found to be a satisfactory description of 
such sub-dominant terms in the description of the elliptical beam transform.
Note however that the larger the beam ellipticity $\chi$, 
the worse this approximation turns out to be.


\subsection{Prescriptions for an accurate analysis}
\label{sec:presc}

As was shown above, the convergence of the perturbative development is
fast enough so that very few terms of the expansion are needed. To
quantify this convergence and to define some useful prescriptions, we
compare it to exact numerical integration and determine the order $n$
of the $b_{\l 0}$ expansion needed in order to obtain an agreement
better than $1\%$ up to $\l_{max} = 5 ~\l_{peak}$, where
$\l_{peak}$ denotes the maximum of the window function as defined in
Eq.(\ref{lpeak}). Some prescriptions are summarized in
table~(\ref{table_n}), where we write this order for two different beams
of full width half
maximum, $\theta_{\rm FWHM} = 5'\displaystyle$ and $10'$ and a set of
reasonable ellipticities.  Even if we see that naturally the greater
the ellipticity, the greater is the required $n$, as a matter of fact,
in most of practical situations (see \S\ref{sec:disc}),
$3$~terms at most are needed.

Note that this criteria is very stringent and if we require, say, 
only a
$2\%$ accuracy at the peak level, only $1$ ellipticity correction is
needed for
$\chi \le 0.3$, \ie $\sigma_a/ \sigma \le 1.2$.

{\footnotesize
\bigskip
\begin{center}
\begin{tabular}{lc|c|c|c|c|c}
\multicolumn{0}{r}{$\chi$} & \multicolumn{1}{l|}{$({\sigma_a/
\sigma})$} & ${\tiny 0.17 (1.1)} $ & ${\tiny0.30 (1.2)}$ &${\tiny0.40
(1.3)}$ & ${\tiny0.49 (1.4)}$ & ${\tiny0.55 (1.5)}$ \\
\hline
\hline
\multicolumn{1}{l|}{$\theta_{{\tiny FWHM}}$} & $\l_{peak}$ & 548 & 531
& 519
& 509  & 502 \\
\multicolumn{1}{l|}{$\quad  10'$ }      & $n$      & 2     &  3   &   4
&
6  & 6 \\
\hline
\multicolumn{1}{l|}{$\theta_{{\tiny FWHM}}$} & $\l_{peak}$  & 1097  &
1063 &
1039 & 1020 & 1005 \\
\multicolumn{1}{l|}{$\quad 5'$}        & $n$      & 2     &  3   &   4
&
6  & 7
\end{tabular}
\end{center}
{\bf Table~I.} 
Required number of terms $n$ in the ellipticity expansion, Eq(\ref{blmn}),  
to achieve a precision greater than $1\%$ up to $\l_{max} = 5 ~\l_{peak}$
for beams of different $\theta_{\rm FWHM}$ and ellipticity $\chi$.
\label{table_n}
\bigskip
}

The numbers presented in this table lead to another
requirement. Indeed, as was discussed in \S~\ref{sec:slight} the
$n$-th order correction to the $m=0$ mode is of the same order
than the leading order contribution to the $m=n$ mode ($m$ even) (see
Eqs.(\ref{blm}) \& (\ref{highm})). 
Thus to be self-consistent, \emph{the highest perturbative 
order, $n$, in the ellipticity
corrections to the circular ($m=0$) mode of the window, $b_{\l\,0}$, 
should match the highest-$m$ mode considered for an accurate 
computation of the full beam transform, $b_{\l\,m}$}.
For example, 
the previous table implies that to handle properly the elliptical
beam effects at a $1\%$ precision till $\l_{max} = 5 ~\l_{peak}$,
\eg for a beam of $\theta_{\rm FWHM}=10^{\prime}$, 
we have to include $m=2$ mode
for $\chi = 0.17$ or $0.30$ , while one has to include $m=2,4$ modes 
for $\chi = 0.40$, and so forth.

\section{Beam Spherical Harmonic Transform: Linear Polarization}
\label{sec:linpol}

The CMB radiation is expected to be linearly polarized as 
caused by Thomson scattering of CMB photons off hot electrons
primarily at the surface of last scattering 
(see \cite{Re68}, \cite{Ba80}, \cite{Ka83}, \cite{BoEf84}) while 
foreground Galactic emission is observed to be linearly polarized
as well (see \eg \cite{OlTe99} for recent reviews and references therein). 
Thus we shall focus here
on the detection of linearly polarized radiation and neglect 
circular polarization in what follows.

The case for a linearly polarized beam with an elliptical shape
can be treated in an analog way to the formalism developed in 
\S\ref{sec:intensity} for the total intensity beam.

Linearly polarized radiation can be conveniently described 
in terms of the so-called Stokes parameters, $\widetilde{Q}$ 
and $\widetilde{U}$ (note that 
we use $\widetilde{X}$ to denote beam parameters,
as opposed to sky parameters, $X$). 
Stokes parameters of a plane wave are related
to the amplitudes of the electric field of the wave in two 
directions orthogonal to the wave propagation direction.

Following standard notation 
(see e.g \cite{St96}, \cite{Se97}, \cite{KaKo97a},
\cite{KaKo97b},\cite{ChFo00}),
the Stokes parameters of the beam are decomposed in the spin-2 spherical
harmonics basis $_{\pm 2}\; Y_{\l \,m}$ as, 
\be
{1 \over \sqrt{2}}(\widetilde{Q} \pm i \widetilde{U}) = 
\sum_{\l m} (b_{\l \,m}^G \mp ib_{\l\,m}^C)\, _{\mp 2}Y_{\l \,m}  
\en
Equivalently, the harmonic transform of a linearly polarized beam 
in terms of the so-called Gradient ``G'' and Curl ``C'' components reads,  
\be
b_{\l \,m}^G \pm ib_{\l \,m}^C = {1 \over \sqrt{2}} \int d\Omega\;
(\widetilde{Q} \mp i \widetilde{U})\, _{\pm
2}Y_{\l \,m}^{\ast}   
\label{blmgc}
\en
from which it follows that
\bea
b_{\l \,m}^G &=& {1 \over 2\sqrt{2}} 
\int d\Omega\; \Bigl[ (\widetilde{Q} - i\widetilde{U}) \, _{2}Y_{\l \,m}^{\ast} +
(\widetilde{Q}+i\widetilde{U}) \, _{-2}Y_{\l \,m}^{\ast} \Bigr]
\nonumber  \\ 
b_{\l \,m}^C &=& {-i \over 2\sqrt{2}} 
\int d\Omega\; 
\Bigl[ (\widetilde{Q} - i\widetilde{U})\, _{2}Y_{\l \,m}^{\ast} -
(\widetilde{Q}+i\widetilde{U}) \, _{-2}Y_{\l \,m}^{\ast} \Bigr]
\label{blmgc1}
\ena
where the above expressions assume that 
the available power to each of the modes (G,C) is
$1/2$ of the total intensity (ie, we assume 
fully polarized detectors, with no sensitivity to circular polarization).
Note that this $G$ (Gradient) and $C$ (Curl) 
components of the linear polarization
are simply linked to the $E$ and $B$ ones, respectively, in the following way, 
\be
b_{\l \,m}^E = -\sqrt{2}\ b_{\l \,m}^G  \quad 
b_{\l \,m}^B = -\sqrt{2}\ b_{\l \,m}^C   \quad ;
\en
See \cite{Za01} for a pedagogical discussion of the $E, B$ polarization modes.
For a pure co-polar beam (ie, for an ideal optical
system and telescope, see \cite{ChFo00}), 
we have
\be
\widetilde{Q}\pm i\widetilde{U} = -B(\theta,\phi) ~e^{\pm 2i\phi}
\label{stokesB}
\en
where $B(\theta,\phi)$ is defined as in Eq.(\ref{bpolar}).
Eq.(\ref{stokesB}) reflects the spin-2 nature of linear polarization
in the $(\theta,\phi)$ basis.

Let us evaluate the harmonic transform of the linearly polarized beam.
Using the parity symmetries for an elliptical beam (see Appendix 2),
\bea
b_{\l \,m}^C &=& i ~b_{\l \,m}^G  \ , \quad 
b_{\l \,-  m}^C = -i ~b_{\l \,-m}^G  
\ena
and the general (intrinsic to the definition of the G,C components) 
parity transformations,
\footnote{Note that these conjugation rules are consistent
with \cite{KaKo97b} \& \cite{NgLi99}, and are inconsistent
by a factor $(-1)^m$ with respect to \cite{Za98}.},  
\bea
b_{\l \,-m}^P &=& b_{\l \,m}^P , \qquad {\rm P=G,C}
\label{bconj}
\ena
one realizes that 
the harmonic transform of linear polarization can be fully determined 
from one of the two components alone, say G. 
Moreover, both negative and positive modes
have the same harmonic transform.
Thus, in what follows, we shall assume $m \ge 0$
without loss of generality.

In full analogy with the total intensity computation 
(see \S\ref{sec:intensity}) we introduce a 
perturbative expansion of the elliptical beam,
\be
b^G_{\l \,m} = \sum_{n} b_{\l \,m}^{G\,(n)} {\chi^n \over n!} =
b_{\l \,m}^{G\,(0)} + b_{\l \,m}^{G\,(1)} \chi + {\cal O} (\chi^2)
\label{blmgpert}
\en
This expansion can be exactly integrated for any order in an analogous 
way to the total intensity case and yields (see Appendix 2 for the 
basic steps of the computation),
\be
b^G_{\l \,m} = ~\sigma^{m-2} ~N^{G}_{\l\,m}
~e^{-z} ~\sum^{\infty}_{\nu=0} \gamma_{\nu, m-2} 
~L^{(m-2)}_{\nu}(z) ~\chi^{\nu+m/2-1} 
\label{blmg}
\en   
where we define 
$N^{G}_{\l\,m} = -{\l}^{2m} ~M_{\l\,m}/(4\sqrt{2}\bar{B_0})$, 
and the coefficients $\gamma_{\nu, m-2}$ are the same than 
those defined for the total intensity 
Eq.(\ref{blmn}). The normalization $M_{\l\,m}$ is given in Appendix 2 
along with the basic notation for the spin-2 harmonics.

Note that, up to the normalization $N^G_{\l\,m}$, 
the linear polarization beam transform Eq.(\ref{blmg}) is formally 
the same than the total intensity one Eq.(\ref{blmn}), with the index
for the m-modes shifted by $m \rightarrow m-2$. 
This shift is introduced by the difference in spin-index $s$ between 
the linearly polarized beam $s=2$ (see Eq.(\ref{stokesB}) above) and 
the total intensity beam $s=0$.

In particular, Eq.(\ref{blmg}) shows 
that {\em the $m=2$ mode dominates the harmonic
transform of the linearly polarized elliptical beam}. 
Note that for a circular Gaussian
beam ($\chi = 0$) {\em only} the $m=2$ mode is non-vanishing.
To leading order
in the ellipticity expansion ($\nu=0$ in Eq.(\ref{blmg})) one finds
that contributions from $m>2$ modes
are sub-dominant,
\be
b^G_{\l \,m} = - \gamma_{0, m-2} \chi^{m/2-1} 
~(\sigma \l)^{m-2} ~{N_{\l\,0} \over 2\sqrt{2}} 
~e^{-\l^2\sigma^2/2} 
\label{highmg} 
\en
with $m=4,6,\ldots$
In particular, one finds the same suppression of higher $m$-modes with respect
to $m=2$, 
in full analogy to the results for the total intensity Eq.(\ref{highm}).
Also, the linear polarization window function peaks at increasingly higher
$\l$-multipoles as one considers higher $m$-modes, as was the case for
the total intensity window.  

The expressions for the first non-zero $m$-contributors 
($m=2$ and $m=4$ modes) 
up to the first ellipticity correction are
\be
b^G_{\l \,2} = -{N_{\l\,0}\over 2\sqrt{2}}
~e^{-\l^2\sigma^2/2} ~\Bigl[1-{\chi\over 4} \l^2\sigma^2 \Bigr] 
+ {\cal O} (\chi^2)
\label{bl01g}
\en
\be
b^G_{\l \,4} =  -{N_{\l\,0}\over 2\sqrt{2}}  ~{\chi \over 8} 
\l^2\sigma^2 ~e^{-\l^2\sigma^2/2} + {\cal O} (\chi^2)
\label{bl21g}
\en
and therefore,
\be
b^G_{\l \,4} =  ~{\chi \over 8} \l^2\sigma^2 ~b^G_{\l \,2} +
{\cal O} (\chi^2)
\label{bl4r}
\en
and the same expressions hold for negative modes, 
as $b^G_{\l\,-m}=b^G_{\l\,m}$.
This is in 
full analogy with the scaling relation between higher $m$-modes
found for the total intensity window Eq.(\ref{bl2r}).

Whenever the beam ellipticity is fairly large, one has to consider
higher-order corrections in the ellipticity to compute accurately the
window function.  
Explicit expresssions for the window function up to {\em second order} in the
ellipticity are given in the Appendix A, Eq(\ref{app:blm2gor}).

\newpage
\section{Full-sky Polarization Correlation Matrix}
\label{sec:spectra}

Linearly polarized radiation is described by
the total intensity $T$ and the Stokes parameters $Q$ and $U$. 
If the CMB polarized radiation is Gaussian distributed,
one needs, a priori, $6$ statistical quantities to characterize 
correlations among them.
It is more convenient to use linear combinations of the Stokes parameters 
with different parity properties, the so-called
$E$ (or Gradient $G$ ) and 
$B$ (or Curl $C$) modes, for which only $4$ correlations are
non-vanishing. 
Namely the correlation between $E$ and $T$
modes and the $3$ autocorrelations.

Following \cite{Za98} we will consider the correlation matrix
$\mathbf{M}$ between $2$ arbitrary measurements in the sky 
\be
\mathbf{M}(\widehat n_1,\widehat n_2) \equiv \left(
\begin{array}{ccc}
<T_1 T_2> & <T_1 Q_2> & 0 \\
<T_1 Q_2> & <Q_1 Q_2> & 0 \\
0         & 0         & <U_1 U_2>
\end{array}
\right)
\en
where $1, 2$ denote the directions $\widehat n_1, \widehat n_2$ in the sky.
The cross terms $<T_1 U_2> = <Q_1 U_2>= 0$ as required by symmetry 
under parity transformations (see \eg \cite{KaKo97b}).

The entries of the correlation matrix are defined as follows:
\bea
<P_1 P_2> &=& ~
<P^{\,\ast}_{eff}(\widehat n_1, \omega_1) P_{eff}(\widehat n_2, \omega_2)>\
; \nonumber
\\
P_{eff}(\widehat n, \omega) &=& ~\int d\Omega ~D(\phi, \theta, \omega) 
~\widetilde{P}^{\,\ast} \, P
\label{polcorr}
\ena
where $P_{eff}$ is the result of convolving the polarized beam 
$\widetilde {\rm P} = \widetilde {\rm T},\widetilde {\rm Q},
\widetilde {\rm U}$  
with the sky,  P $=$ T, Q, U.

In this formalism the ``scanning strategy'' of a given 
experiment is obtained by specifying the Euler angles as a function of
time $t$, $(\phi(t), \theta(t), \omega(t))$, where
$\widehat{n} = \widehat{n}(\theta(t),\phi(t))$
gives the pointing direction of the beam and $\omega(t)$ is the
rotation angle around the pointing direction $\widehat{n}$ which
specifies the orientation of an asymmetric beam (eg, the major axis orientation
for an elliptical Gaussian beam) with respect to a fix reference
orientation (eg, ~$\omega(t=0)$). 

Accordingly, 
the rotation operator $D(\phi, \theta, \omega)$ acts on the beam
so that it takes all possible orientations with respect to a fix
reference frame in the sky \cite{WaGo00}, \cite{ChFo00}.
Simple scanning strategies allow a convenient decomposition of the
rotation matrix $D(\phi(t), \theta(t), \omega(t))$
for the implementation of fast methods to compute the full-sky convolution
\cite{WaGo00}, \cite{ChFo00}.
In what follows we shall suppress the 
time dependence of the Euler angles to simplify notation.

Decomposing the polarization field 
in spin harmonics one finds the following expressions for the
Stokes parameters of the beam 
convolved with the sky (see Eq.(5) of \cite{WaGo00} and 
Eq.(39) of \cite{ChFo00}).
\be
T_{eff}  = \sum [D^{\l}_{m \, M}(\phi, \theta, \omega)]^\ast
~b_{\l\,M}^\ast ~a_{\l\,m}
\en
\be
Q_{eff} = 2
\sum [D^{\l}_{m \, M}(\phi, \theta, \omega)]^\ast
~b^{G\,\ast}_{\l\,M}~a^G_{\l\,m} 
\en
\be
U_{eff} = 2
\sum [D^{\l}_{m \, M}(\phi, \theta, \omega)]^\ast
~b^{G\,\ast}_{\l\,M}~a^C_{\l\,m} 
\en
where we define, 
$a^G_{\l\,m} = (a_{2\, ,\,\l\,m} + a_{-2\, ,\,\l\,m})/2\sqrt{2}$ and
$i a^C_{\l\,m} = (a_{2\, ,\,\l\,m} - a_{-2\, , \,\l\,m})/2\sqrt{2}$.

Note that for the linear polarization parameters $Q, U$,
an overall factor of 2 
accounts for the fact that both G and C modes contribute equally 
to the transform of an elliptical beam.

The polarization correlation matrix can be easily computed making use of
symmetry properties of the rotation matrices and 
the addition theorem of rotations (see Appendix 3),
\be
<T_1 T_2> = 
\sum_{\l} 
~b^{TT}_{\l} C^T_{\l} 
\label{tt}
\en
\be
<T_1 Q_2> = 
2  \sum_{\l}
~b^{TG}_{\l} C^{TG}_{\l} 
\label{tq}
\en
\be
<T_1 U_2> \propto ~C^{TC}_{\l} = 0
\en
\be
<Q_1 U_2> \propto ~C^{GC}_{\l} = 0
\en
\be
<Q_1 Q_2> = 
4 \sum_{\l}
~b^{GG}_{\l} C^G_{\l} 
\label{qq}
\en
\be
<U_1 U_2> = 
4  \sum_{\l}
~b^{GG}_{\l} C^C_{\l} 
\label{uu}
\en
where we have used the fact that $C^{TC}_{\l} = C^{GC}_{\l}  = 0$
as follows from the general property that 
the $T, G$ harmonic coefficients of a field 
transform differently under parity than the $C$ harmonics. 
The power spectra are defined as,
\be
< a^{P \,\ast}_{\l\,m} ~a^{P}_{\l^{\prime} m^{\prime}} > = C^P_{\l}
\delta_{\l\,\l^{\prime}} \delta_{m\,m^{\prime}} \quad , \quad {\rm P =T, G, C}
\en
and we have introduced the ``2-point window functions'',
\be
b^{P\,P^{\prime}}_{\l}  
= \sum_{M\,M^{\prime}} D^{\l}_{M^{\prime} M}
~b^{P \,\ast}_{\l\,M}  ~b^{P^{\prime}}_{\l\,M^{\prime}} \quad ; 
\quad ${\rm P = T, G}$ 
\label{2pblm}
\en
where hereafter we drop the tilde to denote beam quantities,
(\ie we take ${\rm \widetilde{P} \rightarrow P}$) to ease notation.
We remind that the first non-vanishing contributions to
the total intensity $b^T_{\l\,m} = b_{\l\,m}$ enter at $m=0$, while  
linear polarization beams $b^G_{\l\,m}$ have the first non-zero contributor
from $m = 2$.
 
\newpage

The rotation matrix $D^{\l}_{M^{\prime} M}$ above reads (see Eq.(2) in 
\S4.7.1 and Eq.(5) in \S4.7.2 of \cite{VaMo88}) 
\bea
D^{\l}_{M^{\prime}M}(\alpha-\omega_2, \beta, \gamma-\omega_1)
&=& d^{\l}_{M^{\prime} M}(\beta) 
~e^{-i [M^{\prime} (\alpha-\omega_2) + M (\gamma-\omega_1)]} 
\nonumber \\
\label{rotmat}
\ena
where the Euler angles $(\alpha, \beta, \gamma)$ 
of the resulting rotation matrix are (Eq.(6) in \S4.7.2 of {\cite{VaMo88}; 
see also \cite{SoRa01}),
\bea
\cot\alpha &=& \cos\theta_2 \cot(\phi_1-\phi_2) - 
\cot\theta_1\sin\theta_2\csc(\phi_1-\phi_2) \nonumber \\
\cos\beta &=&  \cos\theta_1 \cos\theta_2 +
\sin\theta_1 \sin\theta_2 \csc(\phi_1-\phi_2) \nonumber \\
\cot\gamma &=&  \cos\theta_1 \cot(\phi_1-\phi_2) - 
\cot\theta_2 \sin\theta_2\csc(\phi_1-\phi_2) 
\label{Dangles}
\ena
and the orientation of the beam at pixels $1$ and $2$ is given by 
$\omega_1$ and $\omega_2$.
The {\em irreducible} rotation matrices 
(see Eq.(2.17) of \cite{BrSa62} and Eq.(4) in \S 4.3.1 of
\cite{VaMo88}), read
\bea
d^{\l}_{n m}(\beta) &=& 
\sum_t (-1)^t 
{\Bigl[(\l+n)! (\l-n)!(\l+m)! (\l-m)! \Bigr]^{1/2} \over
t! (\l+n-t)! (\l-m-t)! (t+m-n)!} \nonumber \\
&\times& (\cos\beta/2)^{2\l+n-m-2t} (\sin\beta/2)^{2t+m-n}
\label{dmat}
\ena
where $t$ is summed up for all values which yield non-negative factorials.
These matrices
give the dependence of the polarization correlation functions 
on the separation (or lag) angle in the sky 
$\beta = \widehat n_1 \cdot \widehat n_2$.
Thus, predictions for the polarization 
correlation matrix for cosmological models of the sky signal 
convolved with an elliptical window
assuming a particular scanning strategy 
are given by Eqs(\ref{blm}),(\ref{blmg}),(\ref{Dangles}) \& (\ref{dmat}).

The polarization correlation matrix thus obtained can be used to
compute the likelihood functions and the Fisher information matrix for
a given sky realization of a cosmological model 
convolved with an elliptical beam.

\newpage
\subsection{Slightly Elliptical Beams}

Provided the beam ellipticity is small ($\chi \ll 1$) a first order
ellipticity correction to the circular Gaussian beam 
yields a good approximation to the elliptical beam transform 
(see \S\ref{sec:slight}). Consistently, one can expand
the polarization correlation functions to first order in the 
ellipticity expansion.

For this purpose, first we need to explicit the first terms 
of the 2-point window,
\bea
b^{P P^{\prime}}_{\l} &=&   
D^{\l}_{00} b^P_{\l\,0} b^{P^{\prime}}_{\l\,0}
+D^{\l}_{02} b^P_{\l\,0}b^{P^{\prime}}_{\l\,2} + \cdots 
\label{bttexp}
\ena
where we have taken the real part of the 2-point
function as we want to compute the polarization correlation functions
which are observable and therefore real.
Eq.(\ref{bttexp}) for the case of the total intensity 
($P=T$) is in agreement with Eq.(33) in \cite{SoRa01}.
Introducing the 
expansion of the (1-point) windows $b^P_{\l\,m}$ to first order
in the ellipticity 
Eqs(\ref{bl01}),(\ref{bl21}),(\ref{bl01g}) \& (\ref{bl21g}) 
into Eq.(\ref{bttexp}) one obtains the following expressions
for the correlation functions,
\bea
<T_1 T_2> &=& 
\sum_{\l} C^T_{\l} 
\Bigl[ D^{\l}_{00} + {\chi\over 2}\l^2\sigma^2 
\Bigl(D^{\l}_{02} - D^{\l}_{00} \Bigr) \Bigr] \nonumber \\
&\times& \Bigl({2\l+1 \over 4\pi} \Bigr)~e^{-\l^2\sigma^2}  +{\cal O}(\chi^2)
\label{ttexp}
\ena
\bea
-<T_1 Q_2> &=& \sqrt{2}
\sum_{\l}  C^{TG}_{\l} 
\nonumber \\ 
&\times& \Bigl[ D^{\l}_{02} + {\chi\over 8}\l^2\sigma^2 
\Bigl(D^{\l}_{04}+ D^{\l}_{22} - 4 D^{\l}_{02} \Bigr) \Bigr] 
\nonumber \\ 
&\times& \Bigl({2\l+1 \over 4\pi} \Bigr)
~e^{-\l^2\sigma^2} + {\cal O}(\chi^2) \label{tqexp}
\ena
\bea
<Q_1 Q_2> &=&
\sum_{\l}  C^G_{\l} 
\Bigl[ D^{\l}_{22} + {\chi\over 4}\l^2\sigma^2 
\Bigl(D^{\l}_{24} - 2 D^{\l}_{22} \Bigr)  \Bigr] \nonumber \\
&\times& \Bigl({2\l+1 \over 4\pi}\Bigr)~e^{-\l^2\sigma^2} +{\cal O}(\chi^2) 
\label{qqexp}
\ena
\bea
<U_1 U_2> &=&
\sum_{\l}  C^C_{\l} 
\Bigl[ D^{\l}_{22} + {\chi\over 4}\l^2\sigma^2 
(D^{\l}_{24} - 2 D^{\l}_{22} \Bigr) \Bigr] \nonumber \\
&\times& \Bigl({2\l+1 \over 4\pi} \Bigr)~e^{-\l^2\sigma^2} +{\cal O}(\chi^2)
\label{uuexp}
\ena
where the sum involving the rotation matrices 
$D^{\l}_{M\,M^{\prime}}$ is restricted to $\l \ge M+M^{\prime}$, and   
\bea
D^{\l}_{00} &=& d^{\l}_{00}(\beta) = P_{\l}(\cos\beta)  \nonumber \\
D^{\l}_{02} &=& d^{\l}_{02}(\beta) \Bigl[\cos2\alpha+\cos2\gamma \Bigr] \nonumber \\ 
D^{\l}_{22} &=& d^{\l}_{22}(\beta) \cos2(\alpha+\gamma)  \nonumber \\
&+& (-1)^{\l}d^{\l}_{22}(\pi-\beta)\cos2(\alpha-\gamma)  \nonumber \\
D^{\l}_{04} &=& d^{\l}_{04}(\beta) \Bigl[\cos4\alpha+\cos4\gamma \Bigr]
\nonumber \\ 
D^{\l}_{24} &=& d^{\l}_{24}(\beta) \Bigl[\cos(2\alpha+4\gamma)
+\cos(4\alpha+2\gamma) \Bigr] 
\nonumber \\ 
&+& (-1)^{\l}d^{\l}_{24}(\pi-\beta) \Bigl[ \cos(2\alpha-4\gamma)
+ \cos(4\alpha-2\gamma) \Bigr]  
\label{Dterms}
\ena
where we have to replace $\alpha \rightarrow\alpha -\omega_2, 
\gamma \rightarrow\gamma -\omega_1$ to introduce the beam orientation in the
above equations, as determined by Eq.(\ref{rotmat}).
We have used the symmetry properties of the $d$-rotation matrices 
(see \S4.4 Eq.(1) in \cite{VaMo88}) $d_{M M^{\prime}}(\beta) =
(-1)^{M^{\prime}-M}d_{M M^{\prime}}(\beta), 
d_{M -M^{\prime}}(\beta) =
(-1)^{\l+M}d_{M M^{\prime}}(\pi-\beta)$, and the reality condition
on the correlations functions.
The {\em irreducible} $d_{sm}$-matrices 
can be expressed in terms of Legendre Polynomials
by relating the previous to the spin-$s$ harmonics (see Eqs(3.4), (3.11) in 
\cite{Goetal67}), 
\be
d^{\l}_{sm}(\beta) =  \sqrt{{4\pi}\over {2\l+1}}\,  _{-s}Y_{\l\,m}(\beta, 0)
\en
For $s=0, 2$ one gets (see Eqs(\ref{app:ylm}) \& (\ref{app:2ylm})),
\bea
d^{\l}_{0m}(\beta) = \sqrt{(\l-m)!\over(\l+m)!}~P^m_{\l}(\cos\beta),
\nonumber \\
d^{\l}_{2m}(\beta) = 2 \sqrt{(\l-2)!(\l-m)!\over(\l+2)!(\l+m)!}
\, _{-2}P^m_{\l}(\cos\beta) 
\ena
and $_{-2}P^m_{\l}(\cos\beta)$ are given in Eq.(\ref{app:2plm}).
This explicitly shows that the ellipticity (asymmetry) of the window
function introduces a dependence of the correlation functions
on the scanning strategy, as parameterized by the angles 
$(\alpha, \beta, \gamma)$. 
We stress that the above equations are appropriate for a full-sky analysis,
since the small-angle approximation is only taken for the 
beam geometry (which is of small extent in radians).

In Fig \ref{2pttlin} we display the temperature correlation
function $<TT> \displaystyle$ (we drop sub-indices labeling
sky pixels to ease notation) for a slightly elliptical 
beam Eq.(\ref{tt}).
We assume a flat power spectrum, $C_{\l} = Constant$. This allows us to 
emphasize the effect of the window function irrespective of the
underlying cosmological model assumed. 
For the case shown in Fig \ref{2pttlin}, we assume 
that the telescope scans the sky in ecliptic latitude
(\ie $\alpha = \gamma = 0$) and that the beam hits a given sky pixel
always with the same orientation (\ie we consider correlated pixels 
for beams aligned in the sky) which provides an upper limit
to the effect of ellipticity on the correlation functions.
This is because scanning strategies which observe
a given sky pixel with a different orientation of the beam each time
it scans over it, tend to average out the impact of beam asymmetry
on full-sky estimators.

\begin{figure}[!bh]
\plotone{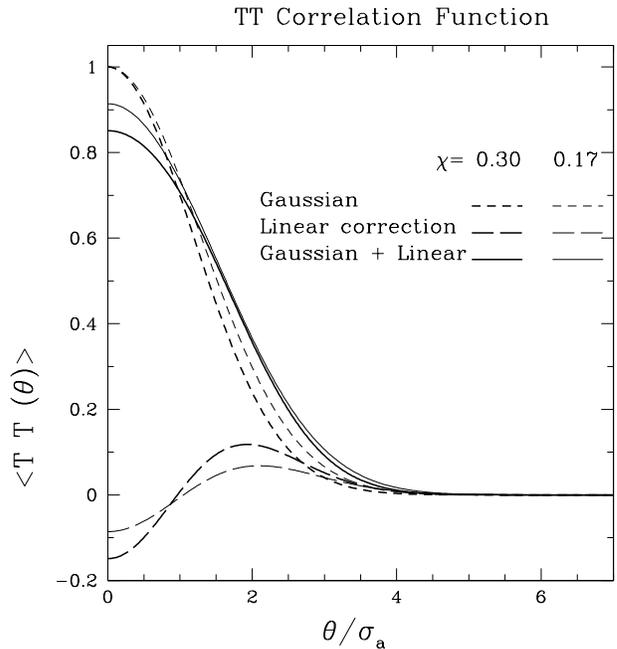}
\caption{Temperature correlation function for a 
slightly elliptical beam. It assumes a flat power spectrum, $C_{\l} =
constant$.
(Short dashed line) 
Correlation function for a circular Gaussian beam. 
(Long dashed) Linear Ellipticity correction
assuming the beam scans the sky in ecliptic latitude and 
a fixed beam orientation in the sky. (Solid) Total correlation
function (Gaussian + linear correction). Thick lines assume $\chi = 0.3 
(\sigma_a/\sigma=1.2$) while thin lines correspond to 
$\chi = 0.17 (\sigma_a/\sigma=1.1$).
\label{2pttlin}}
\end{figure}

\begin{figure}[!bh]
\plotone{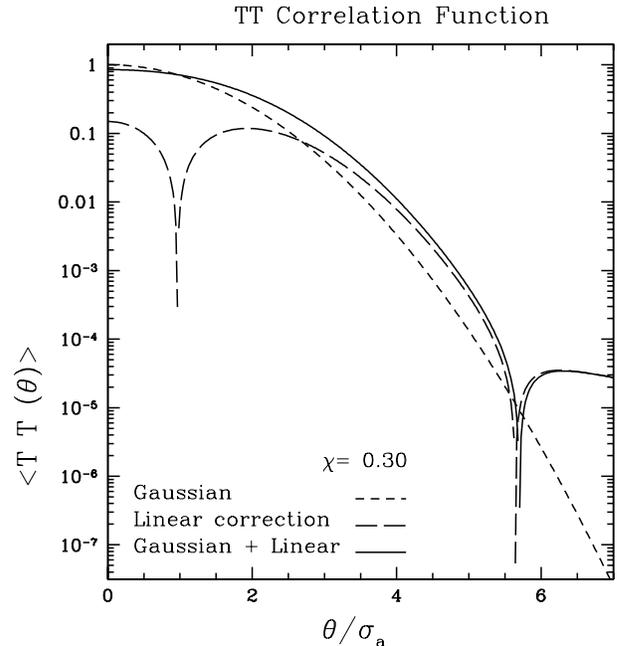}
\caption{Same as Fig \ref{2pttlin} but for the absolute value 
of the correlations in log scale to emphasize
small residual correlations induced by the beam at large angular
separations. 
Only lines for 
an ellipticity of $\chi=0.3$ are shown for clarity. 
\label{2pttlog}}
\end{figure}

\begin{figure}
\plotone{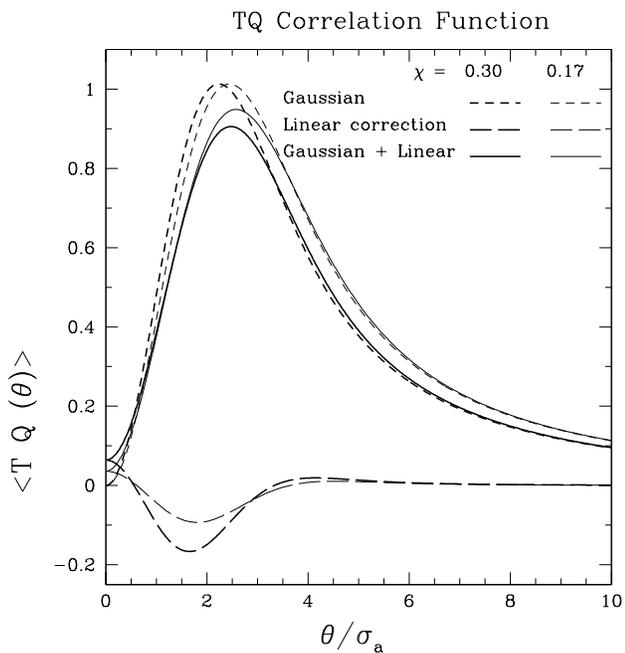}
\caption{Cross-correlation 
Temperature-Polarization. Conventions are the same than in 
Fig \ref{2pttlin}.
\label{2ptqlin}}
\end{figure}

\begin{figure}
\plotone{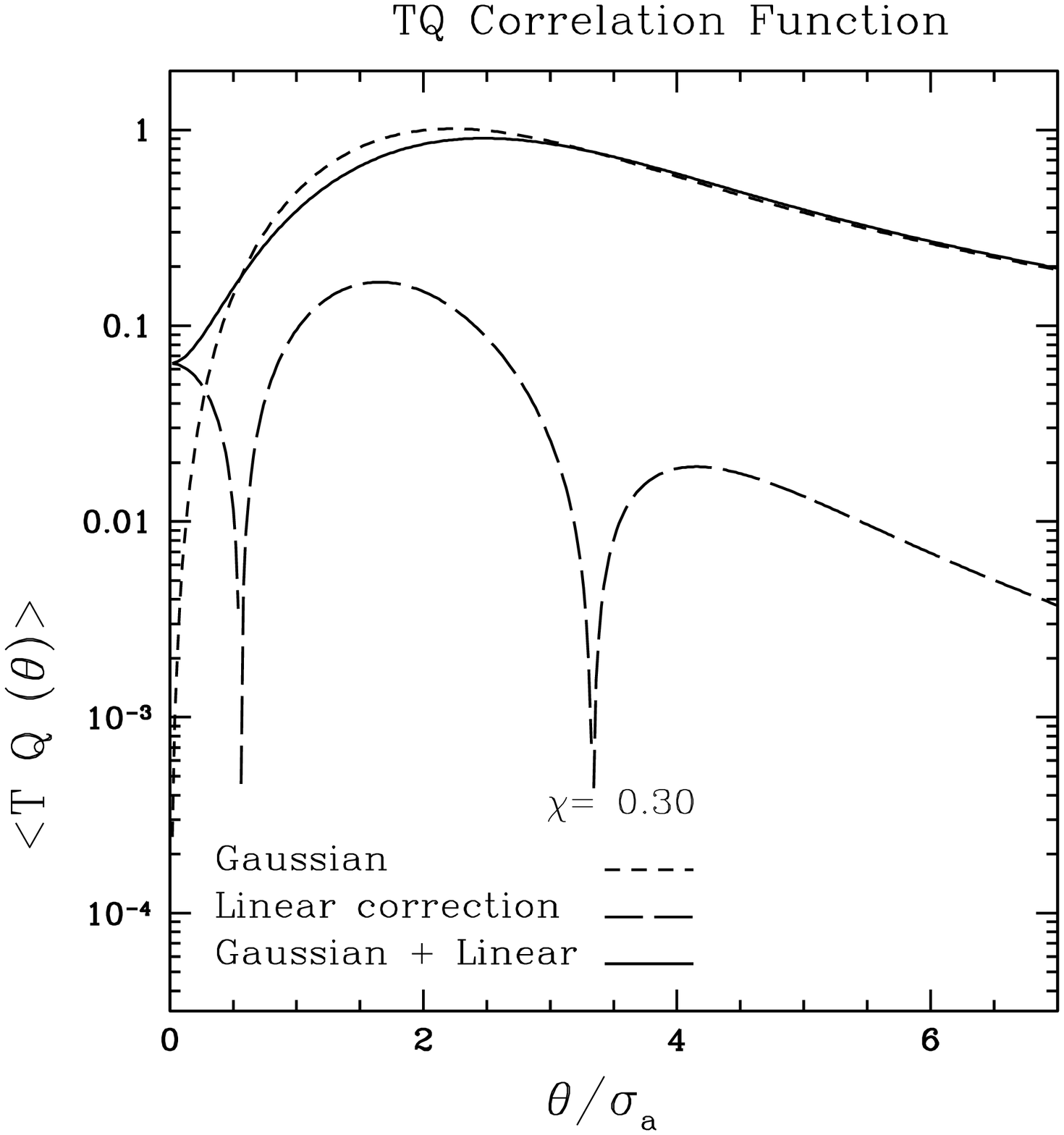}
\caption{Same as Fig \ref{2ptqlin} but in log scale.
\label{2ptqlog}}
\end{figure}

\begin{figure}
\plotone{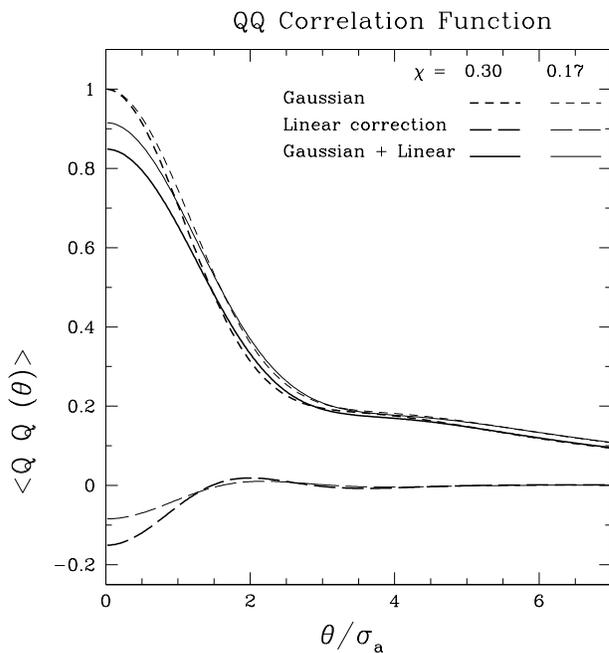}
\caption{
Linear Polarization correlation function in
terms of the Stokes Q parameter. Note that for 
the case shown (a flat power spectrum), $<QQ>=<UU>$. 
Conventions are the same than in Fig \ref{2pttlin}.
\label{2pqqlin}}
\end{figure}

\begin{figure}
\plotone{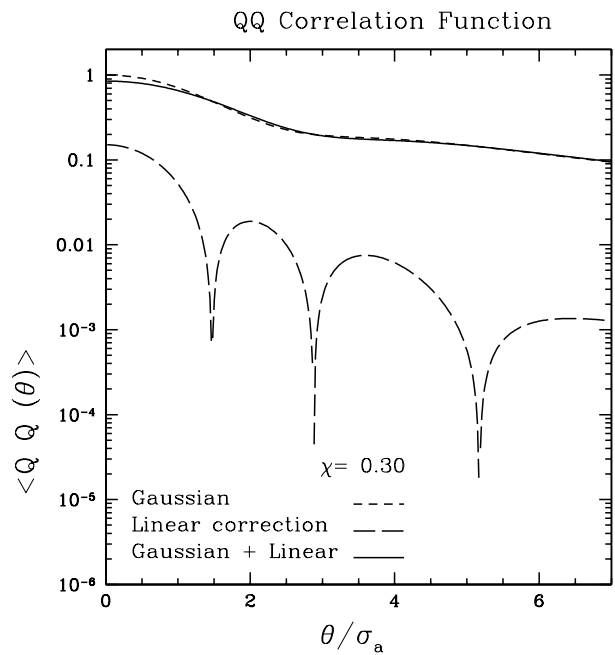}
\caption{Same as Fig \ref{2pqqlin} but in log scale.
\label{2pqqlog}}
\end{figure}

\newpage
As seen in the plots, the linear ellipticity 
correction to the circular Gaussian
window introduces an anti-correlation for pixels separated by $\theta
\le \sigma_a$ (the major axis beam-width). This is because 
pixels within this angular separation are seen as a single {\em
smeared pixel}. Alternatively, for $\theta
\ge \sigma_a$ the ellipticity increases the correlation between 
sky pixels. This correlation peaks at $\theta = 2 \sigma_a$, where
it yields a $20 \%$ correction (for $\chi = 0.3$) 
to the Gaussian correlation function 
and decreases monotonously for larger distances, as expected.
However, a closer look (see Fig \ref{2pttlog}) 
reveals that some small residual anti-correlations induced by the ellipticity
(at the level of $10^{-5}$) 
remain at large distances which might be a reflexion of the
limitations of a linear order analysis. Non-linear terms in the 
ellipticity expansion are expected to cancel out these 
long-range correlations.

Cross-correlations for temperature-polarization and
linear polarization auto-correlation functions are shown
in Figs 
(\ref{2ptqlin}),(\ref{2ptqlog}) \& 
(\ref{2pqqlin}),(\ref{2pqqlog}), respectively.
In particular, we see that the effect of ellipticity is comparable
for the temperature auto-correlation $<TT>$ 
and cross-correlation functions $<TQ>$
(at most a $15 \%$ correction to the Gaussian correlation  
for $\chi=0.3$), while tends to be
less significant for the linear polarization $<QQ>$ (just a few percent
correction).
Notice that the angular scale for the transition between negative and positive
ellipticity-induced
correlations (first and second bumps in the log-scale plots) is shifted in
the correlations involving linearly polarized windows with respect to
the case for the temperature (or total intensity) discussed above
(see Figs (\ref{2pttlin}) \& (\ref{2pttlog})).

\section{Covariance Matrix}
\label{sec:covariance}

In this section we shall discuss the covariance of the 
power spectra for elliptical beams in the presence of
uncorrelated\footnote{the noise is assumed to be uncorrelated between
different pixels and between temperature and linear polarization 
measurements.} noise, 
following the standard formalism developed for circular windows
\cite{Kn95} (see also
\cite{KaKo97a}, \cite{KaKo97b}, 
\cite{Se97}, \cite{ZaSe97}, \cite{Za98} \& \cite{NgLi99}). 
In particular, we shall use this
formalism to estimate error-bars for the power-spectra
for elliptical window functions. 
For this purpose we shall
assume that the circular mode of an elliptical window, which has
a {\it non-Gaussian} profile,
yields an approximately unbiased estimate of the actual error-bars, as
we shall argue below.

The covariance of the temperature power spectrum,
$C_{\l}^T$ can be easily computed for the circular mode ($m=0$)
of the window function \cite{Kn95}
\be
\Delta(C_{\l}^T) = \sqrt{{2 \over (2\l+1)\Delta\l f_{sky}}} 
~\Bigl(C_{\l}^T + w^{-1} ~(b_{\l \,0})^{-2}\Bigl) \,.
\label{clterr}
\en
Similarly for linear polarization, one obtains from  
the lowest $m$-mode contribution ($m=\pm 2$) \cite{KaKo97b}, \cite{ZaSe97},
\be
\Delta(C_{\l}^P) = \sqrt{{2 \over (2\l+1)\Delta\l f_{sky}}} 
~\Bigl(C_{\l}^P + w^{-1} ~(2 \,b^G_{\l \,2})^{-2}\Bigl)
\label{clperr}
\en
where ${\rm P=G, C}$, and for the cross-spectra
\bea
\Delta(C_{\l}^{TG}) &=& \sqrt{{1 \over (2\l+1)\Delta\l f_{sky}}} 
~\left[ (C_{\l}^{TG})^2 \right. \nonumber \\
&+& \left. 
~\Bigl(C_{\l}^{T}  + w^{-1} ~(b^T_{\l \,0})^{-2} \Bigr)
\right. \nonumber \\
&\times& \left. \Bigl(C_{\l}^G + w^{-1} ~(2 \,b^G_{\l \,2})^{-2} \Bigr)
\right]^{1/2}
\label{cltperr}
\ena
where the factor of $2$ in the polarization windows 
accounts for equal contributions from $m= \pm 2$. 
The factors $\Delta \l$ and $f_{sky}$ in the above expressions
account for the binning in ${\l}$-space used 
(we assumed $\Delta \l = 75$ for all experiments)
and the fraction of
the sky observed by the experiment, respectively. 
The weight per solid angle is 
$w\equiv (\sigma^2_{pix}\omega_{pix})^{-1}$ 
while the noise per pixel $\sigma_{pix} = s/\sqrt{t_{pix}}$ depends
on the detector sensitivity $s$ and the observing time per pixel $t_{pix}$.
The pixel solid angle $\omega_{pix} =
\theta_{\rm FWHM}\times\theta_{\rm FWHM}$.
The above expressions for the noise associated to power spectra
estimation, Eqs.(\ref{clterr})-(\ref{cltperr}) assume that all
detectors in the experiment have the same noise properties and
main beam response.

Note that polarization power spectra have twice as much noise
per pixel as the temperature spectrum since only half of the total power is 
available to each polarization mode (G and C). 
This is accounted for through 
the normalization of the window functions (see factor $\sqrt{2}$
in Eq.(\ref{blmgc})).\footnote{Alternatively, one can define
different pixel weights $w$ for temperature and linear polarization,
$w^P = 2 ~w^T$ (see \eg \cite{KaKo97b})}

We stress that the above expressions only include the leading order
in the $m$-mode expansion of the elliptical window. However this
is a good approximation to (\ie, it is the dominant term in) 
the exact window function for elliptical
beams as discussed in \S\ref{sec:slight}. 
In principle, this analysis could be rigorously extended to include higher 
$m$-modes of the window by computing the $a_{{\l}\,m}$'s of the
sky map convolved with the elliptical window, from which the
power spectra of the convolved map and their associated errors
can be calculated (see appendix A.2 in \cite{Hietal01}).
However thorough numerical analyses 
(\cite{WuSt00}; see also \cite{BuMa98}) show that an azimuthally symmetrized
component of the window yields an unbiased estimate of the power
spectrum within a few percent, 
what suggests that non-circular modes of the window function
can be safely neglected, at least for slightly elliptical beams.

Predictions for the theoretical error-bars for the 
temperature power spectrum for current CMB experiments are shown in  
Fig \ref{fig:clterr}. 
Experimental parameters are taken from \cite{Haetal00}, 
\cite{Letal01} (MAXIMA-1), \cite{Netal01}, \cite{Pr01} (Boomerang),
\cite{Betal01} (Archeops) and \cite{Ta00} (Planck). 
The figures used correspond to averages among channels and they
only intend to be illustrative. Note that for Archeops and Planck
the experimental numbers given are just nominal.
It is seen that the pixel noise blows up the
error bars at multipoles $\l \approx 1000$, except for the Planck
satellite experiment. 
Main differences between error forecasts for different 
experiments are due to the sky coverage and
noise per pixel (for a single channel).
It is also observed that the error bars computed for a 
Gaussian window underestimate those of an elliptical window (computed
according to Eq.(\ref{bl01}). However
to first order, the error bars for an elliptical beam 
can be well approximated by 
using an {\em effective Gaussian window} of the form 
$b_{\l 0} = \exp[-{\l}^2\sigma_a\sigma/2]$, where $\sigma_a$ and 
$\sigma$ are the major and minor axis of the ellipses of constant beam
response.

\begin{figure}
\plotone{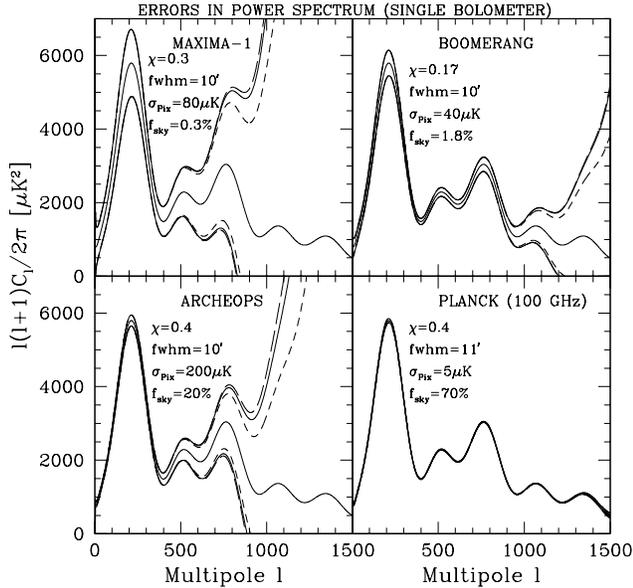}
\caption{Errors in the power spectrum estimation for current and
future experiments.  
It assumes an underlying standard $\Lambda$CDM
model. Pairs of lines (above and under the mean power spectrum) show different
estimates for the theoretical error bars according to different 
choices of the window function: short-dashed lines are predictions for a 
circular Gaussian window of beam-width given by the minor axis of the
ellipse $\sigma$, solid lines correspond to the elliptical window to
first order, Eq.(\ref{bl01}), while long-dashed lines are obtained
from an {\em effective circular Gaussian window} of beam-width
$\sigma_{eff} = \sqrt{\sigma_a\sigma}$, being $\sigma_a$ the major
axis of the ellipse. In all cases, a binning in $\l$-space is used of
width $\Delta \l = 75$.  
\label{fig:clterr}} 
\end{figure}

A detailed analysis of the expected error bars in the 
power spectra estimation including polarization 
for Planck (single 100 GHz channel) is
summarized in Fig \ref{fig:clerrplanck}. As discussed above, the
high sensitivity of Planck allows a clean recovery of the CMB power
spectra up to $\l \simgt 1000$ with a single channel data (except for 
the C-polarization mode, see below). 
In fact, pixel noise starts blowing up the error bars
for the temperature anisotropy power spectrum at $\l \simgt 1500$ (see
upper panel in Fig\ref{fig:clerrplanck}). For the cross-spectrum
temperature-polarization (G-mode) one finds that pixel noise becomes dominant
at $\l \simgt 1000$ (see middle panel) whereas for
polarization (G-mode),
this happens at lower multipoles $\l \approx 1000$ (see lower panel). 
We have checked (although is not shown in Fig \ref{fig:clerrplanck} 
for clarity sake) that for the polarization C-mode error bars become 
pixel-noise dominated at $\l \simlt 500$ as the signal is typically
(\ie for standard CDM models) found at a few percent level of 
that in the G-mode.

Beyond these multipoles (\ie for smaller scales) 
the effect of the ellipticity of the window becomes significant. 
Moreover, using
a circular Gaussian window clearly underestimates error bars
for elliptical beams approximately computed according to
Eqs.(\ref{bl01}) \& (\ref{bl01g}) for the total intensity and linear
polarization windows, respectively.


\begin{figure}
\plotone{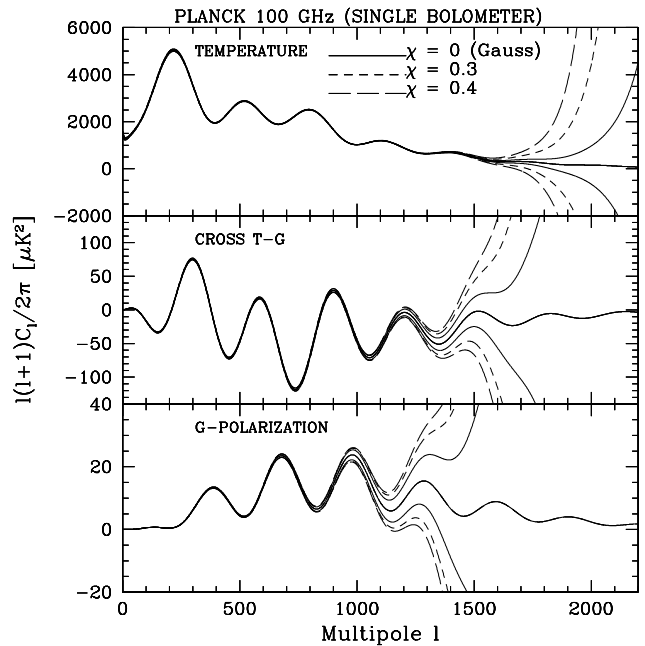}
\caption{Errors in the power spectrum estimation from a single
100 GHz detector of the Planck
satellite. It assumes the same 
experimental parameters than those given in Fig \ref{fig:clterr}
(bottom right panel).
Solid lines show the mean theoretical power pectra and their error bars for
realizations of the sky convolved with a Gaussian beam.
Dashed lines show the analog error bars for the case of an Elliptical beam.
Upper panel displays the temperature anisotropy power spectrum, 
middle panel shows
the cross temperature-polarization (in terms of the G-mode),
while the bottom panel corresponds to the polarization (G-mode).
\label{fig:clerrplanck}} 
\end{figure}

\newpage
\section{Discusion}
\label{sec:disc}

As cosmic microwave background (CMB) experiments image the sky
at finer spatial resolution with higher sensitivity, 
new relevant systematic effects have to be properly taken care of
in the process of data analysis in order to consistently extract 
cosmological information down to the smallest scales probed by the experiment.
The asymmetry of the beam response is becoming an increasingly important issue 
which has been largely neglected until recently in 
CMB studies.

In this paper we have introduced an analytic approach to describe
the effect of beam ellipticity in CMB experiments.
This approach is based on a perturbative expansion around the geometry 
of a circular Gaussian beam, which yields a series expansion of the
elliptical Gasusian beam in powers of the ellipticity parameter.
There are several advantages of introducing a perturbative
approach to discuss beam ellipticity:

\begin{itemize}

\item{
It provides a simple and convenient way of integrating the beam harmonic
transform for the total intensity and linear polarization.}

\item{
In most of current experiments the beam ellipticity is small (we shall refer
to these as ``slightly elliptical beams'' in what follows), i.e, 
the beam fullwidths along the major and minor axis differ by 10-20
$\%$ at most.
This implies that, in practice, the perturbative expansion 
truncated to low orders describes the harmonic transform with 
high accuracy up to very high multipoles.}

\item{
The perturbative expansion allows a simple
qualitative discussion of the
role that different m-modes play in the beam
transform (see e.g, section III). In particular, 
the relative weight of these modes is assessed by working out 
how they depend on experimental parameters (e.g, width and ellipticity
of the beam).
This information cannot be directly extracted 
from a non-perturbative solution.}
 
\item{
The full-sky polarization correlation matrix can be most
simply discussed for the "slighly elliptical beams" for which
deviations from the circular Gaussian beam results
can be explicitly derived.}

\end{itemize}

In particular,
we have obtained analytic solutions for both the total intensity 
(temperature anisotropy) and linear polarization window functions.
The main results are given in \S\ref{sec:intensity}, Eq.(\ref{blm})
\& \S\ref{sec:linpol}, Eq.(\ref{blmg}).

Our findings show that the circular ($m=0$) mode dominates 
the total intensity window function, although the first 
non-circular (higher $m$) modes  
can not be neglected in a consistent analysis.
The reason for the latter is that higher $m$-modes in the beam transform 
can be identified amongst the higher-order corrections in the
ellipticity expansion around the circular Gaussian window.
This provides a simple explanation for previous 
semi-analytic and numerical results in the literature.

For linear polarization, we found that $m=2$ is the dominant mode
but again, higher modes ($m=2,4,\ldots$) must be included 
to compute accurately the window function.

Numerical integration validates our approach and provides 
practical prescriptions for how many terms in the perturbative expansion 
of the circular mode of the window 
have to be taken to achieve a given accuracy. This in turn translates
directly in how many non-circular (higher-$m$) modes 
contribute non-negligibly to the window function of the elliptical
beam (see \S\ref{sec:converge}).

We have implemented our analytic solutions for the elliptical window function  
to derive expressions for the full-sky polarization correlation
functions for elliptical beams (see \S\ref{sec:spectra}).
In particular, we have derived simple analytic expressions for
slightly elliptical beams, taking into account the beam
orientation
and scanning strategy of a given experiment. 
We find that, for simple scanning strategies, 
the ellipticity of the beam induces additional correlations
of the order of $20 \%$ for small angular separations (few
beam-widths) with respect to a circularly symmetric Gaussian beam.

Finally, we have investigated the impact of beam asymmetry in 
error estimation for CMB power spectra in the presence of uncorrelated
noise. We find that error bars for
a circular Gaussian window largely underestimate those of an
elliptical window when the pixel noise becomes dominant.
However, a good approximation to the actual error bars is given
by an {\em effective circular Gaussian} window of beam-width
$\sigma_{eff} = \sqrt{\sigma_a\sigma}$, being $\sigma_a$ and 
$\sigma$ the major and minor axis of the ellipse.
Note that, for slightly elliptical beams ($\chi \rightarrow 0$), 
$\sigma_{eff} \approx  \sigma (1+\chi /4)$ which is approximately the width
of the circular mode of the elliptical window, $b_{\l 0}$, as
discussed in \S\ref{sec:intensity}. This explains why for
quasi-circular windows, $\sigma_{eff} = \sqrt{\sigma_a\sigma}$,
provides an accurate estimate of the power spectra error-bars.

We shall emphasize that, in our approach,
we introduce the experimental beam 
in the {\it time stream}, while the "effective beam" in the 
{\it pixel domain}
is the result of multiple observations of the same sky-pixel 
with different orientations of
the beam (and possibly with a different noise level), 
for general scanning strategies. 
This implies that non-circular modes of the "effective beam"
are expected to cancel out to some extent and therefore
the "effective" circular component of the beam should yield an almost
unbiased estimate of the $C_{\l}$, 
as shown by recent numerical analysis (see Wu et al 2001).
Therefore the nominal ellipticity in the time
domain will be typically larger than the final 
"effective" ellipticity on the map. 
In the discussion of the estimated errors in the 
power-spectrum presented in \S\ref{sec:covariance}, we take the 
``effective'' ellipticity to be the same than the one defined in the 
time stream and thus our estimates must be taken only as upper limits
to the actual effect of window ellipticity.

The issue of beam asymmetry here discussed 
is particularly relevant for future high-resolution and sensitivity
CMB anisotropy experiments, especially those
measuring also polarization, such as the Planck satellite.

In a future work \cite{Foetal01}, we shall validate the
elliptical model for the beam asymmetry presented here in the presence
of other systematic effects (non-elliptic beam distortion/asymmetry,
pointing errors, other sources of noise, etc). Such analysis
will show under which circumstances beam ellipticity is a major systematic
effect in a realistic analysis of a CMB experiment. 
Some recent work along this lines has already been done for the Planck
satellite \cite{Chetal01}, 
although the formalism used is only valid for small-patches of the sky.

\acknowledgments
We would like to thank E. Elizalde, E. Hivon, R. Juszkiewicz, S. Prunet, 
E. Simonneau, and especially R. Stompor and R. Teyssier 
for many useful comments and discussions.
We also want to thank an anonymous referee 
for pointing out some issues which helped us improve the manuscript. 
PF acknowledges a CMBNET Research Fellowship from 
the European Comission. 



\appendix

\section{Perturbative Expansion of the Elliptical Beam Harmonic Transform}

In this appendix we present the key steps for the derivation of the
spherical harmonic transform for the total Intensity beam
Eq.(\ref{blm}). 

In the flat-sky limit ($\theta \rightarrow 0$) the elliptical beam
shape can be expressed in cartesian (x,y),
\bea
B(x, y) & = & B_0(\sigma_a,\sigma_b)\ 
\exp\Bigl[-{x^2 \over {2\sigma^2_a}} - {y^2 \over {2\sigma^2_b}} \Bigr]
\ena 
where we define $\sigma_a$ and $\sigma_b$ as the beam-widths in the
major x and minor y axis, and the normalization is given
by $B_0(\sigma_a,\sigma_b) = 1/(2\pi \sigma_a \sigma_b)$.

For analysis on the sphere, it is more convenient to introduce
(planar) polar coordinates to describe the beam,
$x=\theta\cos\phi$ 
and $y=\theta\sin\phi$, 
\bea
B(\theta, \phi) & = & 
B_0\ \exp\Bigl[-{\theta^2 \over 2\sigma_b^2} f(\phi)\Bigr] 
\label{app:ellibeam}
\ena 
where $f(\phi) \equiv 1 - \chi\cos^2\phi$ describes 
deviations from the circular (or axisymmetric) Gaussian window
and the ellipticity
parameter $\chi \equiv 1- (\sigma_b/ \sigma_a)^2$, is defined within the range $1 \ge \chi \ge 0$.
The circular Gaussian window is thus 
recovered for the limiting case $\chi = 0$.
For the sake of simplicity, we have taken the beam to be
pointing to the north pole of the sphere ($\theta=0$).

The spherical harmonic transform of the total intensity beam
is defined as,
\be
b_{\l \,m} = \int d\Omega ~B(\theta,\phi) ~Y^*_{\l \,m}(\theta,\phi)
\label{app:blm}
\en
where, $d\Omega = d\theta \sin \theta d\phi$, 
and spherical harmonics are defined as,
\be
Y_{\l \,m}(\theta,\phi) = N_{\l \,m} P^m_\l(\cos\theta) ~e^{im\phi} 
\label{app:ylm}
\en
\be
N_{\l \,m} = \sqrt{2\l +1 \over 4\pi} \sqrt{{(\l -m)! \over
(\l+m)!}} \nonumber
\en
where $P^m_\l$ are the Legendre polynomials 
and the spherical harmonics obey the conjugation 
property $Y^*_{\l \,m} = (-1)^m Y_{\l\,-m}$.
Replacing Eq.(\ref{app:ylm}) in
Eq.(\ref{app:blm}) we get,
\bea
b_{\l \,m} &=& (-1)^m N_{\l\,-m} \nonumber \\
&\times& \int^{\pi}_{0} d\theta \,\sin\theta
P^{-m}_\l(\cos\theta) \int^{2\pi}_{0} d\phi B(\theta, \phi) \,e^{im\phi}
\label{app:blm2}
\ena

In the flat-sky limit ($\theta \ll 1$ rad, $\l \gg 1$),   
\be
P^{-m}_\l(\cos\theta) \approx \l^{-m} J_m(\l\,\theta)
\en
where $J_m$ is the m-th order Bessel function of the first kind.
In this limit, the above integral Eq.(\ref{app:blm2}) reads,
\be
b_{\l \,m} = (-1)^m N_{\l\,-m} \l^{-m}
\int^{\pi}_{0} d\theta \,\theta J_m(\l\,\theta)
\int^{2\pi}_{0} d\phi B(\theta, \phi)e^{im\phi}
\label{app:blmflat}
\en
In order to solve this integral analytically, we
introduce a convenient perturbative expansion of the beam 
in real space in powers of the ellipticity parameter $\chi$,
\bea
B(\theta,\phi) &=& B_0 ~B(\theta) 
~\exp \Bigl[\chi ~{\theta^2~\over 2\sigma^2} \cos^2\phi \Bigr]
\nonumber \\
&=& B_0 ~B(\theta) 
~\sum_{n=0}^{\infty} \Bigl({\theta^2~\over 2\sigma^2}\Bigr)^n 
\cos^{2n}\phi ~{\chi^n \over n!}
\label{app:beampt}
\ena 
being $B(\theta) = \exp \Bigl[-\theta^2 /2\sigma^2 \Bigr]$,
which yields an analogous series in harmonic space,
\be
b_{\l \,m} = \sum_{n} b_{\l \,m}^{(n)} {\chi^n \over n!} =
b_{\l \,m}^{(0)} + b_{\l \,m}^{(1)} \chi + {\cal O} (\chi^2)
\label{app:blmpt}
\en
The perturbative expansion Eq.(\ref{app:beampt}) allows to factorize
the 2D-integrals of the beam harmonic transform Eq.(\ref{app:blm2}) in
two 1D-integrals for $\theta$ and $\phi$ respectively. 
Thus the n-th order term of the beam transform can be expressed as
follows:
\be
b_{\l \,m}^{(n)} = (-1)^m N_{\l\,-m} ~I^{(n)}_{\l\,m} ~K_m^{(n)}
\label{app:blmnpt}
\en
with,
\bea
I^{(n)}_{\l\,m} &=& \l^{-m} ~{(2\sigma^2)}^{-n}
\int^{\pi}_{0} d\theta \,\theta^{2n+1} \,J_m(\l\,\theta) \,B(\theta)
\label{app:thetaint}
\ena
\bea
K^{(n)}_{m} &=& 
\int^{2\pi}_{0} d\phi \,\cos^{2n}\phi ~e^{-i 2 m \phi} \ .
\label{app:phint}
\ena

Making use of Eq.(6.631.1) of \cite{GrRy65} and Eqs(13.1.27), (13.6.9)
of \cite{AbSt64} one gets,
\be
I^{(n)}_{\l\,m} = \sigma^{2+m} {(n-m/2)! \over 2^{m/2}} 
~e^{-z} ~L^{(m)}_{n-m/2}(z) 
\label{app:thetaint2}
\en
with $z=\l^2\sigma^2/2$ and,
\be
K_m^{(n)} =
{{2\pi}\over 2^{2n}} ~{2n! \over (n+m/2)! (n-m/2)!} 
\label{app:phint2}
\en
for $m$ even, and $K_m^{(n)} = 0$ for $m$ odd. The fact that 
odd $m$-modes do not contribute to the harmonic transform is due
to the parity symmetries of the ellipse. 
Thus the $n$-th order term of the expansion Eq.(\ref{app:blmnpt}) is given by,
\be
b_{\l \,m}^{(n)} = {2\pi \over 2^{2n+m/2}} ~N_{\l\,-m} ~{{2n!}\over (n+m/2)!} 
~\sigma^{2+m} ~e^{-z} ~L^{(m)}_{n-m/2}(z)
\label{app:blmnpt2}
\en
which replaced in Eq.(\ref{app:blmpt2}) yields the final expression,
\be
b_{\l \,m} = \sigma^{m} ~N^I_{\l\,m}
~e^{-z} ~\sum^{\infty}_{\nu=0} \gamma_{\nu, m} 
~L^{(m)}_{\nu}(z) ~\chi^{\nu+m/2} 
\label{app:blmpt2}
\en  
where we define $N^I_{\l\,m} = N_{\l\,-m}/\bar{B_0}$, 
$\bar{B_0} = B_0/(2\pi\sigma^2)$, and
$\gamma_{\nu, m}=(2\nu+m)!/(2^{2\nu+3m/2}(\nu+m/2)!(\nu+m)!)$. 
The first Laguerre Polynomials are,
\bea
L^{(m)}_0 (z) &=& 1  \quad ; \quad L^{(m)}_1(z) = m + 1 - z \nonumber \\ 
L^{(m)}_2 (z) &=& {1\over 2} \Bigl[(m+1)(m+2) + z (-4 -2 m+z)\Bigr]   \nonumber \\ 
L^{(m)}_3 (z) &=& {1\over 6} \Bigl[(m+1)(m+2)(m+3) \nonumber \\ 
&+& 
z\Bigl(-3(m+2)(m+3) + z (9+3 m -z)\Bigr) \Bigr] 
\label{app:lagpol}
\ena
and higher-orders can be obtained from the recurrence relation
(see Eq.(4.18.1) in \cite{Le72})
\bea
L^{(m)}_n(z) &=& {1\over n} \Bigl[(2n-1-x+m)L^{(m)}_{n-1}(z)-
(n-1+m)L^{(m)}_{n-2}(z) \Bigr] \nonumber \\
\ena
In most practical situations the beam ellipticity is rather small,
$\chi \ll 1$.
In these cases, one only needs to compute the first terms ($2$ or
$3$ terms account for the beam transform up to very large multipoles 
with high accuracy, see 
\S\ref{sec:presc}, Table \ref{table_n})

For example, the beam harmonic transform up to {\em second order} in the  
ellipticity expansion 
has non-vanishing contributions only from the modes 
$m=0, 2$ and $4$, which read,
\bea
b_{\l \,0} &=& N_{\l\,0} ~e^{-\l^2\sigma^2/2} ~\left[1-{\chi\over
4} \l^2\sigma^2  \right. \nonumber \\
&+& \left. {\chi^2\over 4} \left(-\l^2\sigma^2+{3\over 16}\l^4\sigma^4 \right)
\right] \nonumber \\
b_{\l \,2} &=& N_{\l\,0} ~{\chi \over 8} ~\l^2\sigma^2 ~e^{-\l^2\sigma^2/2} 
~\left[1+{\chi}\left(1-{1\over 4}\l^2\sigma^2 \right) \right] 
\nonumber \\
b_{\l \,4} &=& N_{\l\,0} ~{\chi^2 \over 128} ~\l^4\sigma^4 
~e^{-\l^2\sigma^2/2} 
\label{app:blm2or}
\ena
where $N_{\l\,0}=\sqrt{2\l+1/4\pi}$, 
and negative modes (\ie $m=-2,-4$) have to be included as 
they have the same harmonic transform than positive modes, \ie 
$b_{\l\,-m} = b_{\l\,m}$. 
Similar expressions for the harmonic transform to {\em first order} in the
ellipticty are given in \S\ref{sec:slight}, Eqs.(\ref{bl01}) \& (\ref{bl21}).
Note that for a circular Gaussian window, $\chi=0$, one gets as expected,
$b_{\l \,0} = N_{\l\,0} \exp[-\l^2\sigma^2/2]$ and 
$b_{\l \,m} = 0$ for $m \ne 0$.


\section{Perturbative Expansion of the Linearly Polarized 
Elliptical Beam Harmonic Transform}
The aim of this appendix is to provide a detailed derivation of 
the harmonic transform for linearly polarized elliptical beams Eq.(\ref{blmg}).
The spherical harmonic transform of a 
linearly polarized beam can be written in terms of the Stokes
parameters $\widetilde{Q}$ \&  $\widetilde{U}$,
\bea
b_{\l \,m}^G &=& {1 \over 2\sqrt{2}} 
\int d\Omega\; \Bigl[ (\widetilde{Q} - i\widetilde{U}) \, _{2}Y_{\l \,m}^{\ast} +
(\widetilde{Q}+i\widetilde{U}) \, _{-2}Y_{\l \,m}^{\ast} \Bigr]
\nonumber  \\ 
b_{\l \,m}^C &=& {-i \over 2\sqrt{2}} 
\int d\Omega\; 
\Bigl[ (\widetilde{Q} - i\widetilde{U})\, _{2}Y_{\l \,m}^{\ast} -
(\widetilde{Q}+i\widetilde{U}) \, _{-2}Y_{\l \,m}^{\ast} \Bigr]
\label{app:blmgc1}
\ena
where we define the spin-2 spherical harmonics as
\footnote{Note that 
$_{\pm 2}Y_{\l \,m} = W_{\l \,m} \pm i X_{\l \,m}$, 
according to the notation used by \cite{KaKo97b}.}
\bea
_{\pm 2}Y_{\l \,m} & = & M_{\l \,m}\ _{\pm 2}P^m_\l(\cos\theta)~e^{im\phi}
\label{app:2ylm}
\ena
where
\be
M_{\l \,m} = 2 \sqrt{{(l-2)! \over (l+2)!}} N_{\l \,m}
\en
and we define a generalization of the Legendre polynomials for 
spin-2 harmonics,
\footnote{The $_{\pm 2}P^m_\l$ polynomials 
are simply related to the $G_{\l \,m}$ 
in \cite{St96}, \cite{KaKo97b}, \cite{Za98}: 
$_{\pm 2}P^m_\l (x) = G^{+}_{\l \,m} (x) \mp G^{-}_{\l \,m} (x)$.} 
\bea
_{\pm 2}P^m_\l(\cos\theta) 
&=& - \Bigl({\l-m^2 \over \sin^2\theta} + 
{1\over 2} \l(\l-1)\Bigr)P_{\l}^m(\cos \theta) 
\nonumber \\ 
&+& (\l+m){\cos \theta \over\sin^2\theta} P_{\l -1}^m(\cos\theta) 
\nonumber \\ 
&\mp& {m \over \sin^2\theta}\Bigl( (\l-1)\cos\theta
P_{\l}^m(\cos\theta) 
\nonumber \\ 
&-& (\l+m)P_{\l -1}^m (\cos \theta) \Bigr)
\label{app:2plm}
\ena

The above quantities obey the following parity conditions:
\bea
_{\pm 2}P^{-m}_{\l} &=& 
(-1)^m {(\l-m)! \over (\l+m)!} \ _{\pm 2}P^m_{\l}
\label{app:2plmcon}
\ena
\bea
M_{\l \,-m} &=&  {(\l+m)! \over (\l-m)!} M_{\l \,m}
\label{app:mlmcon}
\ena
which imply that,
\be 
\ _{\pm 2}Y_{\l \,m}^\ast = \ _{\mp 2}Y_{\l \,m} ~e^{-2 i m \phi} 
\en
what allows us to recast Eq.(\ref{app:blmgc1})
in a more convenient way,
\bea
b_{\l \,m}^G &=& {1 \over 2\sqrt{2}} 
\int d\Omega \Bigl[ (\widetilde{Q} - i\widetilde{U}) \, _{2}Y_{\l \,m} +
(\widetilde{Q}+i\widetilde{U}) \, _{-2}Y_{\l \,m} \Bigr] e^{-2 i m\phi}  
\nonumber \\
b_{\l \,m}^C &=& {-i \over 2\sqrt{2}} 
\int d\Omega \Bigl[ (\widetilde{Q} - i\widetilde{U}) \, _{2}Y_{\l \,m} -
(\widetilde{Q}+i\widetilde{U}) \, _{-2}Y_{\l \,m} \Bigr] e^{-2 i m
\phi} \nonumber \\   
\label{app:blmgc2}
\ena
For a pure co-polar beam (ie, for an ideal optical
system and telescope, see \cite{ChFo00}), 
we have
\be
\widetilde{Q}\pm i\widetilde{U} = -B(\theta,\phi) ~e^{\pm 2i\phi}
\label{app:stokesB}
\en
where $B(\theta,\phi)$ is defined in Eq.(\ref{app:ellibeam}).
We have assumed that the beam response is measured  
in the co- and cross-polar basis defined on the sphere, 
${\bf \sigma_{co}}$ and ${\bf \sigma_{cross}}$, 
according to Ludwig's 3rd definition
\cite{Lu73},
\bea
{\bf \sigma_{co}} &=& \sin\phi ~{\bf \sigma_{\theta}} + 
\cos\phi ~{\bf \sigma_{\phi}} \nonumber \\
{\bf \sigma_{cross}} &=& \cos\phi ~{\bf \sigma_{\theta}} - 
\sin\phi ~{\bf \sigma_{\phi}}
\label{app:lud3}
\ena
where ${\bf \sigma_{\theta}}$ and ${\bf \sigma_{\phi}}$ are the usual
spherical polar basis.
Such co- and cross- polarization basis, Eq.(\ref{app:lud3}), is obtained by
parallel-transporting 
the local cartesian basis defined at the north pole, 
${\bf \sigma_{x}}$ and ${\bf \sigma_{y}}$, along great circles through the
poles of the sphere (see \eg \cite{ChFo00} for a discussion).  

Replacing Eq.(\ref{app:stokesB}) into Eq.(\ref{app:blmgc2}), one sees that
the first term in Eq.(\ref{app:blmgc2}) is non-vanishing only for negative 
$m$-modes while the second term is non-zero for positive $m$-modes
alone. 
What is more, the parity properties of the $G$ and $C$ modes,
\bea
b_{\l \,m}^C &=& i ~b_{\l \,m}^G  \ , \quad 
b_{\l \,-  m}^C = -i ~b_{\l \,-m}^G  \nonumber \\
b_{\l \,-m}^P &=& b_{\l \,m}^P , \qquad {\rm P=G,C}
\label{app:bconj}
\ena 
imply that 
the harmonic transform of linear polarization can be fully determined 
from one of the two components alone, say G. 
Moreover, both negative and positive modes
have the same harmonic transform.
Thus we shall assume $m>0$ below with no loss of generality.
In this case the harmonic transform of the $G$-mode is simply given by,
\be
b_{\l \,m}^G = {M_{\l \,m} \over 2\sqrt{2}} 
\int d\Omega\; 
B(\theta,\phi) \, _{-2}P^m_{\l}(\cos\theta) ~e^{-i(m-2)\phi}  
\label{app:blmg3}
\en
In the flat-sky limit ($\theta \ll 1$ rad, $\l \gg 1$)
\footnote{This corrects the expression for the 
small-angle limit in  \cite{St96}: 
the pre-factor $\l^{m+2}$ in Eq.(\ref{app:sa2pm}) corrects the pre-factor
$\l^{6-m}$ in Eq.(4.32) of \cite{St96}.},   
\be
_{-2}P^{m}_\l(\cos\theta) \approx {1\over 2} (-1)^m 
\l^{m+2} J_{m-2}(\l\,\theta)
\label{app:sa2pm}
\en
and thus,
\be
b_{\l \,m}^G = (-1)^m {\l^{m+2} \,M_{\l \,m} \over 4\sqrt{2}} 
\int d\Omega\; 
B(\theta,\phi) \, J_{m-2}(\l\,\theta)  ~e^{-i(m-2)\phi}  
\label{app:blmg4}
\en

Introducing the ellipticity expansion Eq.(\ref{app:ellibeam}), one
can solve the integral to any perturbative order,
\be
b^{G \,(n)}_{\l \,m} = {1 \over 2\sqrt{2}} M_{\l\,m} 
~\bar{I}^{(n)}_{\l\,m} ~\bar{K}_m^{(n)}
\label{app:blmgpt}
\en
with,
\bea
\bar{I}^{(n)}_{\l\,m} &=& {\l^{m+2} \over 2} ~{(2\sigma^2)}^{-n}
\int^{\pi}_{0} d\theta \,\theta^{2n+1} \,J_{m-2}(\l\,\theta) \,B(\theta)
\ena
\bea
\bar{K}^{(n)}_{m} &=& 
\int^{2\pi}_{0} d\phi \,\cos^{2n}\phi ~e^{-i(m-2)\phi} \ .
\label{app:barkint}
\ena
Noting that the above integrals are basically the same than those for the
total intensity beam Eqs(\ref{app:thetaint}) \& (\ref{app:phint}),
but replacing $m$ by $m-2$ everywhere, 
they can be integrated in the same way,
\be
\bar{I}^{(n)}_{\l\,m} = \sigma^{m} {(n-m/2+1)! \over 2^{m/2}} 
~e^{-z} ~L^{(m-2)}_{n-m/2+1}(z) 
\label{app:thetaintg}
\en
with $z=\l^2\sigma^2/2$ and,
\be
\bar{K}_m^{(n)} =
{{2\pi}\over 2^{2n}} ~{2n! \over (n+m/2-1)! (n-m/2+1)!} 
\label{app:phintg}
\en
for even modes $m \ge 2$, 
and $\bar{K}_m^{(n)} = 0$ for $m$ odd. 
Therefore the $n$-th order term Eq.(\ref{app:blmgpt}) in the beam
expansion is given by,
\be
b_{\l \,m}^{G (n)} = 
{2\pi \over 2^{2n+m/2}} M_{\l\,m} {{2n!}\over (n+m/2-1)!} 
\sigma^{m} ~e^{-z} ~L^{(m-2)}_{n-m/2+1}(z)
\label{app:blmngpt2}
\en
which introduced in Eq.(\ref{app:blmgpt}) finally gives,
\be
b^G_{\l \,m} = \sigma^{m-2} ~N^G_{\l\,m}
~e^{-z} \sum^{\infty}_{\nu=0} \gamma_{\nu, m-2} 
~L^{(m-2)}_{\nu}(z) ~\chi^{\nu+m/2-1} 
\label{app:blmgpt2}
\en  
where we define 
$N^{G}_{\l\,m} = -{\l}^{2m} ~M_{\l\,m}/(4\sqrt{2}\bar{B_0})$, 
and the coefficients $\gamma_{\nu, m-2}$ are the same than 
those defined for the total intensity 
Eq.(\ref{app:blmpt2}), except for the subindex which is $m-2$ here instead 
of $m$ there.

For most of actual experimental beams
the ellipticity is rather small,
$\chi \ll 1$.
As discussed in Appendix A, 
a {\em second order} analysis of the beam ellipticity is already 
very accurate to very large multipoles as comparison with numerical
integration shows
(see \S\ref{sec:presc}, Table \ref{table_n} for specific prescriptions
depending on experimental parameters).
Thus, expanding the beam harmonic transform to {\em second order} in
$\chi$ one gets 
non-vanishing contributions only from the modes 
$m=2, 4$ and $6$,
\bea
b^G_{\l \,2} &=& -{N_{\l\,0}\over 2\sqrt{2}} 
~e^{-\l^2\sigma^2/2} ~\left[1-{\chi\over
4} \l^2\sigma^2  \right. \nonumber \\
&+& \left. {\chi^2\over 4} \left(-\l^2\sigma^2+{3\over 16}\l^4\sigma^4 \right)
\right] \nonumber \\
b^G_{\l \,4} &=& -{N_{\l\,0}\over 2\sqrt{2}} 
~{\chi \over 8} ~\l^2\sigma^2 ~e^{-\l^2\sigma^2/2} 
~\left[1+{\chi}\left(1-{1\over 4}\l^2\sigma^2 \right) \right] 
\nonumber \\
b^G_{\l \,6} &=& -{N_{\l\,0}\over 2\sqrt{2}}
~{\chi^2 \over 128} ~\l^4\sigma^4 
~e^{-\l^2\sigma^2/2} 
\label{app:blm2gor}
\ena
where $N_{\l\,0}=\sqrt{2\l+1/4\pi}$, 
and negative modes (\ie $m=-2,-4,-6$) have to be included as 
they have the same harmonic transform than positive modes, \ie 
$b^G_{\l\,-m} = b^G_{\l\,m}$. 
Analogous expressions for the harmonic transform to {\em first order} in the
ellipticty are given in \S\ref{sec:linpol}, Eqs.(\ref{bl01g}) \& (\ref{bl21g}).
Note that for a circular Gaussian window, $\chi=0$, one gets,
$b^G_{\l \,\pm 2} = -({N_{\l\,0}/2\sqrt{2}}) 
\exp[-\l^2\sigma^2/2]$ and $b^G_{\l \,m} = 0$ for $|m| > 2$.
Note that, as argued above (see paragraph under Eq.(\ref{app:barkint})), 
the linearly polarized beam transform, Eq.(\ref{app:blm2gor}), 
can be straightforwardly obtained from 
the total intensity beam transform, Eq.(\ref{app:blm2or}), by
replacing in the latter $m$ by $m-2$, and including a multiplicative 
normalizing factor of $-1/(2\sqrt{2})$ appropriate for linear
polarization modes, see Eq.(\ref{app:blmgc1}).



\end{document}